\documentclass[11pt]{article}
\usepackage{graphicx}
\usepackage{epsfig}
\usepackage{epsf,amsfonts,amssymb}
\newcommand{\be}{\begin{equation}}
\newcommand{\ee}{\end{equation}}
\newcommand{\bea}{\begin{eqnarray}}
\newcommand{\eea}{\end{eqnarray}}
\newcommand{\ba}{\begin{eqnarray}}
\newcommand{\ea}{\end{eqnarray}}
\newcommand{\nn}{\nonumber \\}

\newcommand{\beq}{\begin{equation}}
\newcommand{\eeq}{\end{equation}}
\newcommand{\beqa}{\begin{eqnarray}}
\newcommand{\eeqa}{\end{eqnarray}}
\newcommand{\beqar}{\begin{eqnarray*}}
\newcommand{\eeqar}{\end{eqnarray*}}

\newcommand{\reef}[1]{(\ref{#1})}

\newcommand{\eg}{{\it e.g.,}\ }
\newcommand{\ie}{{\it i.e.,}\ }

\def\ben{\begin{equation}}
\def\een{\end{equation}}

  \let\n=\nu   
     
\let\C=\Chi

\def\nn{\nonumber} \def\bd{\begin{document}} \def\ed{\end{document}}
\def\ds{\documentstyle} \let\fr=\frac \let\bl=\bigl \let\br=\bigr
\let\Br=\Bigr \let\Bl=\Bigl
\let\bm=\bibitem
\let\na=\nabla
\let\pa=\partial \let\ov=\overline
\def\ba{\begin{array}}
\def\ea{\end{array}}
\def\ft#1#2{{\textstyle{{\scriptstyle #1}\over {\scriptstyle #2}}}}
\def\fft#1#2{{#1 \over #2}}
\def\del{\partial}
\def\vp{\varphi}
\def\sst#1{{\scriptscriptstyle #1}}
\def\oneone{\rlap 1\mkern4mu{\rm l}}
\def\td{\tilde}
\def\wtd{\widetilde}
\def\ie{\rm i.e.\ }
\def\dalemb#1#2{{\vbox{\hrule height .#2pt
        \hbox{\vrule width.#2pt height#1pt \kern#1pt
                \vrule width.#2pt}
        \hrule height.#2pt}}}
\def\square{\mathord{\dalemb{6.8}{7}\hbox{\hskip1pt}}}
\newcommand{\ho}[1]{$\, ^{#1}$}
\newcommand{\hoch}[1]{$\, ^{#1}$}
\newcommand{\lra}{\longrightarrow}
\newcommand{\Lra}{\Leftrightarrow}
\newcommand{\bp}{\tilde \beta^\prime}
\newcommand{\Tr}{{\rm Tr} }
\def\0{{\sst{(0)}}}
\def\1{{\sst{(1)}}}
\def\2{{\sst{(2)}}}
\def\3{{\sst{(3)}}}
\def\4{{\sst{(4)}}}
\def\5{{\sst{(5)}}}
\def\6{{\sst{(6)}}}
\def\7{{\sst{(7)}}}
\def\8{{\sst{(8)}}}
\def\n{{\sst{(n)}}}
\def\cA{{{\cal A}}}
\def\cB{{{\cal B}}}
\def\cF{{{\cal F}}}
\def\cH{{{\cal H}}}
\def\tV{\widetilde V}
\def\tW{\widetilde W}
\def\tH{\widetilde H}
\def\tE{\widetilde E}
\def\tF{\widetilde F}
\def\tA{\widetilde A}
\def\im{{i}}
\def\tY{{{\wtd Y}}}
\def\ep{{\epsilon}}
\def\vep{{\varepsilon}}
\def\R{\rlap{\rm I}\mkern3mu{\rm R}}
\def\bD{{{\bar D}}}

\def\R{\rlap{\rm I}\mkern3mu{\rm R}}
\def\bD{{{\bar D}}}
\def\R{{{\Bbb R}}}
\def\C{{{\Bbb C}}}
\def\H{{{\Bbb H}}}
\def\CP{{{\Bbb C}{\Bbb P}}}
\def\RP{{{\Bbb R}{\Bbb P}}}
\def\Z{{{\Bbb Z}}}
\def\bA{{{\Bbb A}}}
\def\bB{{{\Bbb B}}}
\def\bC{{{\Bbb C}}}
\def\bD{{{\Bbb D}}}
\def\bE{{{\Bbb E}}}
\def\bZ{{{\Bbb Z}}}
\def\Re{{{\frak{Re}}}}
\def\Im{{{\frak{Im}}}}
\def\cosec{{\,\hbox{cosec}\,}}
\def\Gm{{\Gamma_{\!\! -}}}
\def\Gp{{\Gamma_{\!\! +}}}
\def\stan{{standard }}
\def\nonstan{{supernumerary}}

\begin{document}

\setlength{\unitlength}{1mm}

\thispagestyle{empty}
 \vspace*{2.5cm}

\begin{center}
{\bf \Large Euclidean analysis of the entropy functional formalism}\\

\vspace*{2cm}

{\bf \'Oscar J.~C.~Dias,}$^1\,$ {\bf Pedro J. Silva,}$^{2}\,$

\vspace*{0.5cm}

{\it $^1\,$Departament de
F\'{\i}sica Fonamental, Universitat de Barcelona,\\
Av. Diagonal 647,
E-08028 Barcelona, Spain,}\\[.3em]
{\it $^2\,$Institut de Ci\`encies de l'Espai (IEEC-CSIC) and\\
Institut de F\'{\i}sica d'Altes Energies (IFAE),\\
E-08193 Bellaterra (Barcelona), Spain}\\[.3em]

\vspace*{0.3cm} {\tt odias@ub.edu, psilva@ifae.es}

\vspace*{2cm}

\vspace{.8cm} {\bf ABSTRACT}
\end{center}

The attractor mechanism implies that the supersymmetric black hole
near horizon solution is defined only in terms of the conserved
charges and is therefore independent of asymptotic moduli. Starting
only with the {\it near horizon} geometry, Sen's entropy functional
formalism computes the entropy of an extreme black hole by means of
a Legendre transformation where the electric fields are defined as
conjugated variables to the electric charges. However, traditional
Euclidean methods require the knowledge of the {\it full} geometry
to compute the black hole thermodynamic quantities. We establish the
connection between the entropy functional formalism and the standard
Euclidean formalism taken at zero temperature. We find that Sen's
entropy function $f$ (on-shell) matches the zero temperature limit
of the Euclidean action. Moreover, Sen's near horizon angular and
electric fields agree with the chemical potentials that are defined
from the zero-temperature limit of the Euclidean formalism.

\noindent

\vfill \setcounter{page}{0} \setcounter{footnote}{0}
\newpage

\tableofcontents

\setcounter{equation}{0}

\setcounter{equation}{0}\section{\label{sec:intro}Introduction} 

Black holes (BH) are one of most interesting laboratories we have to
investigate quantum gravity effects. Due to their thermodynamic
behavior these objects have been associated to ensembles of
microstates in the fundamental quantum gravity theory where ideally,
quantum statistical analysis should account for all the BH
coarse-grained thermodynamical behavior. In particular, many
important insights in the classical and quantum structure of BH have
been obtained studying supersymmetric configurations in string
theory. Supersymmetric BH have many important properties that turn
out to be crucial to obtain all the new results. Basically,
supersymmetry triggers a number of non-renormalization mechanisms
that protect tree level calculations from higher order loop
corrections. Moreover, this kind of behavior has also been found in
some non-supersymmetric extreme solutions.

\subsection{Attractor mechanism and entropy functional formalism} 

In this context we have the so called attractor mechanism
\cite{attractor}. It was originally thought in the context of four
dimensional $N=2$ supergravity, where  we have that the values of
the scalar fields at the horizon are given by the values of the BH
conserved charges and are independent of the asymptotic values of
the scalars at infinity. For these BH (and others) it has been
checked that the Bekenstein-Hawking entropy agrees with the
microscopic counting of the associated D-brane system. Not only in
the supergravity approximation, but also after higher derivative
corrections are added to the generalized prepotential
\cite{attractor2}. These results motivated a conjecture where the BH
partition function equals the squared of the associated topological
string partition function {\it i.e.}, $Z_{BH}=|Z_{Top}|^2$
\cite{oguri}. Lately, the attractor mechanism has been extended to
other directions, and applied to several gauged and ungauged
supergravities (see, \eg \cite{li,Guica:2005ig,Morales}).

Importantly, the attractor mechanism has provided a new way to
calculate the BH entropy. In a series of articles
\cite{Sen:2005wa,Sen:2005iz,Astefanesei:2006dd}, Sen recovered the
entropy of  $D$-dimensional BPS BH using only the near horizon part
of the geometry. Basically, in this regime the solution adopts the
form $AdS_2 \otimes S^{D-2}$ \footnote{The analysis of the near
horizon geometry has been applied to more general BH that define
squashed $AdS_2\otimes S^{D-2}$ geometries like in
\cite{Astefanesei:2006dd,li}.} plus some electric and magnetics
fields. The entropy $S$ is obtained by introducing a function $f$ as
the integral of the corresponding supergravity Lagrangian over the
$S^{D-2}$. More concretely, an entropy function is defined as $2\pi$
times the Legendre transform of $f$ with respect to the electric
fields $e_i$. Then, an extremization procedure fixes the on-shell
BPS values of the different fields of the solution and in particular
determines the BPS value of the entropy,
 \bea S_{bps}=2\pi\left( e_i {\partial f\over
\partial e_i} -f \right)_{bps}\,.\label{ae}\eea
Note that in the above definition the different near horizon
electric fields take the role of ``conjugated chemical potentials"
to the BH charges. This formalism has also been extended to extreme
non-BPS BH.

The attractor mechanism, both for asymptotically AdS or flat BH,
implies that in the near horizon geometry we have a dual CFT theory
where the microscopic structure can be studied. We expect that not
only the entropy but {\it all} the statistic properties of such
supergravity systems should be described in terms of their dual CFT
states.

\subsection{Zero temperature limit and chemical potentials} 

Supersymmetric BH in asymptotically AdS spaces have also been
studied using the AdS/CFT correspondence \cite{K,Simon,SQSR1,SQSR2}.
For the $AdS_5$ case we still do not have a CFT microscopic
derivation of its entropy that reproduces the supergravity result.
Nevertheless, in \cite{SQSR1,SQSR2} it was showed that the phase
space of this supersymmetric sector can be scanned in both sides of
the correspondence showing a rich structure with phase transitions
and Hagedorn alike behavior\footnote{These $T=0$ phase transitions
were analyzed both in the strong and weak coupling regimes.
Remarkably, it was found that their properties resemble the
well-known finite temperature phase transitions, where the
Hawking-Page phase transition in the strong coupling corresponds to
the deconfinement/confinement transition at weak coupling
\cite{K,SQSR2}.}. In fact, observables in both dual pictures agree
up to numerical factors, a very non-trivial result since the CFT
calculations are performed at zero coupling only\footnote{In
\cite{K} the CFT partition function was calculated at zero coupling.
Also, an index was considered to count supersymmetric states but
unfortunately it turns out to be blind to the BH sector.}. In order
to study the full statistical properties (so that we could in
principle do more than just account for the entropy), in
\cite{SQSR1,SQSR2} it was found how to define the different chemical
potentials $\mu_i$ that control the supersymmetric BH partition
function in the grand canonical ensemble. The basic input comes from
the thermodynamics of the dual CFT theory, where the BPS partition
function is obtained from the finite temperature one, by sending the
temperature to zero. This also sends the several chemical potentials
to their BPS values. The associated dual limiting procedure  in the
supergravity regime corresponds also to send  the temperature to
zero. Done carefully, this defines the supergravity chemical
potentials that are dual to the the CFT ones and, more generally,
the statistical mechanics of {\it supersymmetric} BH that is free of
divergencies. These chemical potentials are the next to leading
order terms of the zero temperature expansion of the horizon angular
velocities and electric potentials. The resulting supergravity
partition function is given, as expected, by the exponentiation of
the regularized Euclidean action $I$ evaluated at the BH solution.
In this paper we call ``Euclidean zero-temperature formalism" to the
zero-temperature limit in the supergravity system that determines
the Euclidean action, entropy and the chemical potentials. After
some algebra we arrive to the supersymmetric quantum statistical
relation (SQSR) \cite{Gibbons:2004ai} where the Euclidean action $I$
can be rewritten as the Legendre transform of the entropy $S$ with
respect to the different supersymmetric chemical potentials $\mu_i$,
\begin{equation}
I_{bps}=\mu_i\,q^i_{bps}- S_{bps}\,, \label{sqsr}
 \end{equation}
where $q^i_{bps}$'s represent the conserved BH charges conjugated to
the $\mu_i$'s (later, we will use the notation
$q^i\equiv\{Q^i,J^i\}$ and $\mu^i\equiv\{\phi^i,\omega^i\}$). As
said above, these supergravity chemical potentials are closely
related to the dual CFT chemical potentials. Therefore, they provide
a very clear picture of the BPS BH as dual to a supersymmetric CFT
in the grand canonical ensemble. This approach also defines the
finite supersymmetric Euclidean action (\ref{sqsr}), and in fact
allows to study the statistical mechanics of BPS black holes. A
similar analysis can be done for extreme non-BPS systems.

\subsection{Entropy functional formalism from an Euclidean perspective} 

Sen's entropy functional formalism is formulated only with the
knowledge of the near horizon geometry. But, since it computes the
BH entropy, which is a thermodynamic quantity, it should be possible
to understand it starting from a traditional thermodynamical
Euclidean analysis of the black hole system.

In fact, the strong resemblance between equations (\ref{ae}) and
(\ref{sqsr}) is evident. In other words,  it would be strange if
\emph{string theory produces two unrelated functions in the same
supergravity regime that calculate the BH entropy}. Looking into
both definitions with more care, we find that the entropy is defined
as the Legendre transform of the BH charges in the saddle point
approximation of the supergravity theory. Nevertheless, in
(\ref{ae}) the vacuum solution is just the near horizon geometry
with conjugated potentials related to the electric fields, and $f$
is the on-shell Lagrangian over only $S^{D-2}$. Instead, in
(\ref{sqsr}), the vacuum is the entire BH solution; the conjugated
potentials are associated to gauge potentials rather than field
strengths; and $I$ is the on-shell full Euclidean action. The main
goal of this paper is to understand the connection between these two
approaches.

One of the key points of our analysis relies in the natural
splitting of the Euclidean action into two parts corresponding
basically to: i) the near horizon part of space, and ii) the
asymptotic region. Then we find that in the extremal cases (without
ergoregion), the asymptotic part vanishes, and the near horizon part
reduces to Sen's function $2\pi f$. Also, the conjugated chemical
potentials found in both methods agree, due to an argument that
relates differences of gauge potentials produced by variations of
near-BPS parameters with variations of the potential on the radial
coordinate.

\subsection{Main results and structure of the paper} 

As stated above, the main goal of this article is to provide a
bridge between Sen's entropy functional formalism and standard
Euclidean analysis of the thermodynamics of a black hole system.
While doing so, we also find that the supergravity conjugated
potentials defined in Sen's formalism map into chemical potentials
of the dual CFT.

We obtain a unifying picture where:

\verb 1) We are able to recover the entropy function of Sen from the
zero temperature limit of the usual BH thermodynamics and the
statistical mechanics definitions of the dual CFT theory. The
supergravity and their dual CFT chemical potentials are identified
with the surviving Sen's near horizon electric and angular fields.
The Euclidean action is identified with Sen's function $2\pi f$.

\verb 2) As a byproduct of the above analysis we have understood how
to calculate the BPS chemical potentials that control the
statistical properties of the BH using only the BPS regime, {\it
i.e.}, without needing the knowledge of the non-BPS geometry. The
CFT chemical potentials are dual to the supergravity ones.
Traditionally, to compute the latter we have to start with the
non-BPS solution and send the temperature to zero to find the next
to leading order terms in the horizon angular velocities and
electric potentials expansions that give the chemical potentials.
This requires the knowledge of the non-BPS geometry. Unfortunately,
sometimes this is not available and we only know the BPS solution.
But, from item 1) we know that the near horizon fields, that Sen
computes with the single knowledge of the BPS near horizon solution,
give us the supergravity chemical potentials. So now we can compute
the supergravity chemical potentials of any BPS BH solution,
regardless of its embedding into a family of non-BPS solutions,
while still keeping the relation with the dual CFT.

\verb 3) It is known that the attractor mechanism seems to work also
for non-supersymmetric but extremal BH \footnote{See
\cite{Astefanesei:2006dd,Dabholkar:2006tb} and references there
in.}. We have tested the Euclidean zero temperature formalism for
many of these BH, always finding a well defined limit and agreement
with Sen's results for extremal non-BPS BH\footnote{Actually at the
level of two derivative theory, Euclidean $T=0$ formalism is well
defined only for BH with no ergoregion. For BH with ergoregion we
have an ill-defined limit, that nevertheless allows to define the
entropy and all chemical potentials. This is telling us that these
geometries are not fully protected from string corrections. The same
caveats and conclusions are also obtained using Sen's approach, and
this is related to the fact that for these BH the attractor
mechanism is only partial since there is dependance on the
asymptotic data \cite{Astefanesei:2006dd}.}. This is a non-trivial
fact since there is no supersymmetry protecting the limit.
Therefore, in general, the supergravity regime should not give the
correct statistical relations. We interpret this result as another
confirmation that there is a protecting mechanism for extremal
non-supersymmetric BH.

The plan of the paper is the following. In section \ref{sec:Sen} we
review Sen's entropy functional approach using the D1-D5-P system as
an illuminating example. In the beginning of section
\ref{sec:QSRflat} we review the main ideas and results of the
Euclidean zero temperature formalism for BH in the AdS/CFT
framework. Then, we apply this formalism to the most general
rotating D1-D5-P system. We analyze  the connection between the
entropy functional and Euclidean formalisms in section
\ref{sec:Map}, identifying how and why both prescriptions are
equivalent. In section \ref{sec:Extreme} we discuss the application
of the Euclidean $T=0$ formalism to extreme non-BPS BH and again
find agreement with Sen's results. Section \ref{sec:Discussion} is
devoted to a short discussion on the results and possible future
avenues to follow. In Appendix \ref{sec:d1d5p} we review the D1-D5-P
BH solution in detail, including its thermodynamics. In Appendix
\ref{sec:several}, we write the chemical potentials and Euclidean
action for some other BH systems not considered in the main body of
the text. We consider the four charged system of type IIA
supergravity, and the Kerr-Newman BH. We confirm that for these BH
the relation established in section \ref{sec:Map} between the
entropy functional and Euclidean formalisms holds. This agreement
also extends to AdS black holes as is explicitly confirmed in the
context of 5$D$ gauged supergravity in \cite{Silva:2007tw}.

\textsl{Note}: While we where proof-reading this article, the paper
\cite{Suryanarayana:2007rk} appeared in the arXives. It contains
relevant discussions and results connected to our work, regarding
Sen's approach and Wald's method for AdS BH.

\setcounter{equation}{0}\section{Entropy functional formalism
revisited}
 \label{sec:Sen}

As we pointed out in the introduction, Sen developed a simple method
-- the entropy functional formalism -- to compute the entropy of
supersymmetric BH in supergravity \cite{Sen:2005wa}. Lately, this
approach has been applied to rotating BH in gauged and ungauged
supergravity (see, \eg \cite{Astefanesei:2006dd,Morales}). Here, we
will review some of the key aspects of this formalism that we will
use latter. We just need to address non-rotating cases, but we will
comeback to rotating attractors at the end of this section, for
completeness.

Sen's entropy functional formalism assumes that: (i) we start with a
Lagrangian $\mathcal{L}$ with gravity plus some field strengths and
uncharge massless scalar fields; and (ii) due to the attractor
mechanism the near horizon geometry of a $D$-dimensional BH is set
to be of the form $AdS_2\otimes S^{D-2}$. From the above input data,
the general form of the near horizon BH solution is
\bea &&ds^2=v_1\left(-\rho^2d\tau+{d\rho^2\over \rho^2}\right)+v_2d\Omega_{D-2}^2\,,\nn\\
&&F^{(i)}_{\rho \tau}=e_i\,,\quad\quad\quad
H^{(a)}=p_a\epsilon_{D-2}\,,\nn\\
&&\phi_s=u_s\,, \label{geralNHsen} \eea
where $\epsilon_{D-2}$ is the unit-volume form of $S^{D-2}$, and
$(e_i,p_a)$ are respectively the electric fields and the magnetic
charges of the BH. Note that $(\vec u,\vec v, \vec e, \vec p)$ are
arbitrary constants up to now and therefore the solution is
off-shell. Next, it is defined the following function
\bea f(\vec u,\vec v,\vec e,\vec p)=\int_{S^{D-2}}\sqrt{-
g}\mathcal{L}\,, \label{geralfsen}  \eea
where $\mathcal{L}$ is the string frame Lagrangian of the theory
(see, \eg (\ref{SIIbstring})). After minimizing $f(\vec u,\vec
v,\vec e,\vec p)$ with respect to $(\vec u,\vec v)$ we obtain the
exact supersymmetric near horizon BH solution in terms of $(\vec e,
\vec p)$. In fact, the field equations are reproduced by this
minimization procedure. Furhermore, minimization with respect to
$\vec e$ gives the electric charges $\vec q$. Explicitly,  the
on-shell values of $\vec u,\vec v,\vec e$ that specify
(\ref{geralNHsen}) for a given theory described by (\ref{geralfsen})
are found through the relations,
\bea {\partial f\over \partial u_s}=0\,,\quad\quad {\partial f\over
\partial v_j}=0\,, \quad\quad {\partial f\over \partial e_i}=q_i\,.
\eea
Then, using Wald formalism \cite{Wald}, Sen derived that the entropy
$S$ of the corresponding BH is given by $2\pi$ times the Legendre
transform of $f$,
\bea S=2\pi\left(e_i{\partial f\over \partial e_i}-f \right)\,.
\label{sen-entropy}\eea
Finally notice that the minimization procedure, can be taken only
after $S$ is defined. In this form $S$ is really an entropy function
of $(\vec u,\vec v,\vec q,\vec p)$, that after minimization equals
the BH entropy as a function of $(\vec q,\vec p)$ only.

In the rest of this section we will discuss the above formalism in a
specific theory. We consider the  D1-D5-P supersymmetric solution of
ten-dimensional type II$B$ supergravity, discussed in the previous
section, as the main example (this case was first analyzed in
\cite{Ghodsi:2006cd}, at the level of supergravity and for its
higher order corrections). Our aim is to highlight the details of
the application of Sen's formalism to this solution. This will
provide a solid background to compare, in section \ref{sec:Map},
Sen's formalism with the Euclidean one developed in section
\ref{sec:QSRflat}.

From Appendix \ref{sec:d1d5pBH} we know that the supersymmetric
D1-D5-P metric, the RR two-form $C_{(2)}$ and the dilaton $\Psi$ are
given by \footnote{This is the string frame version of
(\ref{3charge}), and (\ref{dilaton}) and (\ref{rr}) with
$a_1=a_2=0$.}
\bea &&ds^2={1\over \sqrt{H_1H_5}}[-dt^2+dy^2+{Q_p^{bps}\over
r^2}(dt-dy)^2]\,+
\sqrt{H_1H_5}(dr^2+r^2d\Omega_3^2)+\sqrt{H_1/H_5}\sum_{i=1}^4 dz_i^2\,, \nn\\
&&C_{(2)}=-{Q_1^{bps}\over r^2H_1}\,dt\wedge dy \,-\,Q_5^{bps}
\cos^2\theta d\phi\wedge d\psi\,,\quad\quad e^{2\Psi}={H_1\over
H_5}\,. \label{norotmetric} \eea
where $H_1=(1+{Q_1^{bps}\over r^2})$, $H_5=(1+{Q_5^{bps}\over r^2})$
and $(Q_1^{bps},Q_5^{bps},Q_p^{bps})$ are the D1,D5,P charges,
respectively. Then, it is easy to take the near horizon limit to
obtain,
\bea &&ds^2={\sqrt{Q_1^{bps}Q_5^{bps}}\over4}\left(-\rho^2
d\tau^2+{d\rho^2\over\rho^2}\right)\,+\,\sqrt{Q_1^{bps}Q_5^{bps}}\,d\Omega_3^2\,\nn\\
&&\quad\quad\quad\quad\quad\quad\quad\quad
 +{ Q_p^{bps}\over \sqrt{Q_1^{bps}Q_5^{bps}}}
\left(dz+{\sqrt{Q_1^{bps}Q_5^{bps}}\over2\sqrt{Q_p^{bps}}}\rho
d\tau\right)^2\,
+\sqrt{Q_1^{bps}/Q_5^{bps}}\sum_{i=1}^4 dz_i^2\,, \nn\\
&&F_{(3)}={1\over 2}\sqrt{Q_5^{bps}Q_p^{bps}\over Q_1^{bps}}\,
d\rho\wedge d\tau\wedge dz\, +\,2Q_5^{bps}
\epsilon_3\,,\quad\quad\quad\quad\quad e^{2\Psi}={Q_1^{bps}\over
Q_5^{bps}}\,, \label{NHsen0} \eea
where we used
\bea
 \tau=\frac{2}{\sqrt{Q_1^{bps}Q_5^{bps}Q_p^{bps}}}\,t \,,\qquad \rho=r^2\,,\qquad
 z=y-t\,.
\label{coordtransf} \eea

Note that, alternatively, all the information encoded in the near
horizon structure (\ref{NHsen0}) could be extracted without knowing
the full geometry, using Sen's approach. Its application starts by
assuming that the near horizon metric is given in terms of the
unknowns $(\vec v,\vec u,\vec e,\vec p)$ as follows,
\bea &&ds^2=v_1\left(-\rho^2 d\tau^2 + {d\rho^2\over\rho^2}\right) +
v_2 d\Omega_3^2 + u_1\left(dz+e_2\rho dt \right)^2 +
u_2\sum_{i=1}^4dz_i^2\,, \nn\\
&&F_{(3)}=e_1\,d\rho\wedge d\tau\wedge dz + 2Q_5^{bps} \epsilon_3
\,,\quad\quad e^{2\Phi}=u_3^2\,. \label{NHsen} \eea
Having the Lagrangian (\ref{SIIbstring}) of type II$B$ at hand, one
now follows the steps summarized in
(\ref{geralfsen})-(\ref{sen-entropy}) to find the on-shell
expressions for $(\vec v,\vec u,\vec e)$. From (\ref{NHsen}) one has
$\sqrt{-\tilde{g}}=u_1^{1/2}u_2^2 v_1 v_2^{3/2}
\sin\theta\cos\theta$, $F^{(3)}_{\rho\tau z}=e_1$ and
$F^{(3)}_{\theta\psi\phi}=2Q_5^{bps}\sin\theta\cos\theta$. The
entropy function, $S(\vec u,\vec v,\vec q,\vec
p)=2\pi[q_1e_1+q_2e_2-f(\vec u,\vec v,\vec e,\vec p)]$ is then
\bea S(\vec u,\vec v,\vec q, \vec p)= 2\pi{\biggl\{ }q_1e_1+q_2e_2
-\frac{1}{4}u_1^{1/2}u_2^2 v_1 v_2^{3/2}\left[ u_3^2 \left(
\frac{u_1e_2^2}{v_1^2} +\frac{12}{v_2}-\frac{4}{v_1} \right)+
\frac{e_1^2}{u_1v_1^2} - \frac{4p^2}{v_2^3} \right] {\biggr\} } \,.
\nn  \label{def:S0}
 \eea
Minimizing this entropy function with respect to $\vec u,\vec v,\vec
e$ one finds the on-shell attractor values,
\bea &&\vec v = \left(\frac{1}{4} \sqrt{Q_1^{bps}Q_5^{bps}},
\sqrt{Q_1^{bps}Q_5^{bps}}\right)\,,\quad\quad \vec u =
\left({Q_p^{bps}\over \sqrt{Q_1^{bps}Q_5^{bps}}} ,
\sqrt{Q_1^{bps}\over Q_5^{bps}} ,
\sqrt{Q_5^{bps}\over Q_1^{bps}}\right)\,, \nn\\
&&\vec e = \left( {1\over 2}\sqrt{Q_5^{bps}Q_p^{bps}\over
Q_1^{bps}},{1\over 2}\sqrt{Q_1^{bps}Q_5^{bps}\over Q_p^{bps}}
\right)\,,\quad\quad \vec q=
\left(Q_1^{bps},Q_p^{bps}\right)\,,\quad\quad p=Q_5^{bps}\,,\nn \eea
One also finds that $f(\vec q,\vec p)=0$ on-shell. Plugging this
information into the entropy function $S(\vec u,\vec v,\vec q, \vec
p)$ we get
\bea S(\vec q,\vec p) &=& 2\pi\left[ q_1e_1+ q_2 e_2-f \, \right]_{on-shell} \nn\\
       &=& 2\pi\sqrt{Q_1^{bps}Q_5^{bps}Q_p^{bps}}\,, \label{Ssen1} \eea
that is the well known result for this BH.

It will be relevant for section \ref{sec:Map} to stress that the
above analysis can be carried on in the case where the magnetic
field is replaced by its dual electric field. This electric field
comes from the RR seven-form field strength $F_{(7)}$, Poincar\'e
dual of the magnetic part of $F_{(3)}$,
\bea F_{(7)}={2Q_5^{bps}\over r^3H_5^2}\,dr\wedge dt\wedge dy \wedge
dz^1\wedge dz^2\wedge dz^3\wedge dz^4\,. \eea
In the near horizon limit, {\it i.e.}, after taking the change of
coordinates (\ref{coordtransf}), $F_{(7)}$ reduces to
\bea F_{(7)}={1\over 2}\sqrt{Q_1^{bps}Q_p^{bps}\over
Q_5^{bps}}\,d\rho\wedge d\tau\wedge dz \wedge dz^1\wedge dz^2\wedge
dz^3\wedge dz^4\,. \eea
In the next lines we want to recover this near horizon attractor
value for $F_{(7)}$, without making use of the near horizon limit of
the full geometry, {\it i.e.}, using instead a Sen-like approach.

 In this pure electric case, we first notice that there is an extra
pair of conjugated variables $(e_3,q_3)$ and second, that $f$ should
be now calculated on a modified Lagrangian with the $F_{(7)}$ RR
field strength appropriately added. This is an effective
``democratic" Lagrangian supplemented by duality constraints imposed
by hand \footnote{We should emphasize that the introduction of a RR
$p$-form field strength with $p>5$ doubles the number of degrees of
freedom. To get the right equations of motion from \reef{Sdual} we
must then introduce by hand duality constraints relating the lower-
and higher-rank RR potentials. We ask the reader to see
\cite{Bergshoeff:2001pv} for further details.}. The motivation,
limitations and formulation of this effective Lagrangian are
presented in detail in \cite{Bergshoeff:2001pv}. In this context,
the string frame Lagrangian (\ref{SIIbstring}) of the D1-D5-P system
takes the form,
\bea
 \mathcal{L}=\frac{1}{16\pi G_{10}} \left[e^{-2\Psi}\left(
  \tilde{R}-4\partial_{\mu}\Psi \partial^{\mu}\Psi\right)- \frac{1}{2\cdot 3!}
F_{(3)}^2 - \frac{1}{2\cdot 7!} F_{(7)}^2 \right] \,,
 \label{Sdual}
  \eea
where the {\it magnetic} part of the original $F_{(3)}$ field is now
encoded in the $F_{(7)}$ contribution. The D1-branes and D5-branes
source the electric $F_{(3)}$ and $F_{(7)}$ fields, respectively.

In the entropy function formalism, the function $f(\vec u,\vec
v,\vec e)$ is obtained by evaluating  action \reef{Sdual} at the
horizon, {\it i.e.}, by integrating along the $S^8$ sphere. We use
the near-horizon fields (\ref{NHsen}). So, the metric determinant is
$\sqrt{-\tilde{g}}=u_1^{1/2}u_2^2 v_1 v_2^{3/2}\sin\theta
\cos\theta$, $F^{(3)}_{\rho\tau z}=e_1$ and $F^{(7)}_{\rho\tau z
z_1\cdots z_4}=e_3$. The entropy function, $S(\vec u,\vec v,\vec
q)=2\pi[q_1e_1+q_2e_2+q_3e_3-f(\vec u,\vec v,\vec e)]$ is then
\bea S(\vec u,\vec v,\vec q)= 2\pi{\biggl\{ }q_1e_1+q_2e_2+q_3e_3
-\frac{1}{4}u_1^{1/2}u_2^2 v_1 v_2^{3/2}\left[ u_3^2 \left(
\frac{u_1e_2^2}{v_1^2} +\frac{12}{v_2}-\frac{4}{v_1} \right)+
\frac{e_1^2}{u_1v_1^2} + \frac{e_3^2}{u_1v_1^2 u_2^4} \right]
{\biggr\} } \,. \nn \label{def:S}
 \eea
Minimizing this entropy function with respect to $\vec u,\vec v,\vec
e$ one finds the on-shell attractor values
\bea &&\vec v = \left(\frac{1}{4}
\sqrt{Q_1^{bps}Q_5^{bps}},\sqrt{Q_1^{bps}Q_5^{bps}}\right)\,,\quad\quad
\vec u = \left({Q_p^{bps}\over \sqrt{Q_1^{bps}Q_5^{bps}}} ,
\sqrt{Q_1^{bps}\over Q_5^{bps}} ,
\sqrt{Q_5^{bps}\over Q_1^{bps}}\, \right)\,, \nn\\
&&\vec e = \left( {1\over 2}\sqrt{Q_5^{bps}Q_p^{bps}\over
Q_1^{bps}},{1\over 2}\sqrt{Q_1^{bps}Q_5^{bps}\over Q_p^{bps}} ,
{1\over 2}\sqrt{Q_1^{bps}Q_p^{bps}\over Q_5^{bps}} \right)\,,\qquad
\vec q= \left(Q_1^{bps},Q_p^{bps},Q_5^{bps}\right)\,, \nn\\
&&\label{sen} \eea
which are used to obtain the on-shell function:  $f(\vec q)={1\over
2}\sqrt{Q_1^{bps}Q_5^{bps}Q_p^{bps}}$. Then,  use of equation
(\ref{sen}) yields the on-shell entropy value,
\bea S(\vec q) &=& 2\pi \left[ q_1e_1 +  q_2e_2 + q_3e_3- f \, \right]_{on-shell} \nn\\
&=& 2\pi\sqrt{Q_1^{bps}Q_5^{bps}Q_p^{bps}}\,, \label{Ssen} \eea
that is, in this dual computation we indeed recover the value
(\ref{Ssen1}).

As commented in the introduction, the above approach was generalized
to rotating BH in ungauged and gauged supergravities
\cite{Astefanesei:2006dd,Morales}. At the level of two derivative
Lagrangian, rotating BH in ungauged supergravity have their near
horizon geometry fully determined by the entropy functional only if
they have no ergoregion. However, BH with ergoregion show only
partial attractor mechanism, since their entropy functional has flat
directions \cite{Astefanesei:2006dd,Astefanesei:2006sy}. In this
case, minimization does not fix the value of all quantities in the
near horizon geometry. There is some surviving dependance on the
asymptotic value of the scalars, although it fixes the form of
entropy itself and the electric and angular fields.

Generalization to gauged supergravities includes AdS BH into the
discussion. The resulting picture is basically the same, where care
has to be taken when evaluating $f$ due to Chern-Simon terms in the
Lagrangian (see \cite{Morales} for details). In these cases, the
attractor mechanism is related to a non trivial flow between fixed
points at both boundaries of spacetime, the horizon AdS and the
asymptotic AdS at infinity.

\setcounter{equation}{0}\section{Euclidean zero-temperature
formalism: BPS black holes}
 \label{sec:QSRflat}

In \cite{SQSR1,SQSR2} the ``thermodynamics" or better ``the
statistical mechanics" of supersymmetric solitons in gauged
supergravity was studied in detail using an extension of standard
Euclidean thermodynamical methods to zero temperature systems. We
call this approach the Euclidean zero-temperature formalism. BPS BH
can be studied as dual configurations of supersymmetric ensembles at
zero temperature {\it but} non-zero chemical potentials in the dual
CFT. These potentials control the expectation value of the
conjugated conserved charges carried by the BH, like \eg angular
momenta and electric charge.

In these articles, the two main ideas are: \verb"First", there is a
supersymmetric field theory dual to the supergravity theory.
\verb"Second", in this dual field theory the grand canonical
partition function over a given supersymmetric sector can be
obtained as the zero temperature limit of the general grand
canonical partition function at finite temperature. This limit also
fixes the values of several chemical potentials of the system.

To make things more clear, recall that all supersymmetric states in
a field theory saturate a BPS inequality that translates into a
series of constraints between the different physical charges. For
definiteness, let us consider a simple case where the BPS bound
corresponds to the constraint\footnote{This type of BPS bound
appears in two dimensional supersymmetric models like, {\it e.g.},
the effective theory of $1/2$ BPS chiral primaries of $N=4$ SYM in
$R\otimes S^3$ (see
\cite{Corley:2001zk,Berenstein:2004kk,Caldarelli:2004ig}).}: $E=J$.
Then, defining the left and right variables $E^\pm=
\hbox{$1\over2$}(E_\nu\pm J_\nu)$, $\beta_\pm=\beta(1 \pm \Omega)$
the grand canonical partition function is given by
\bea Z_{(\beta,\Omega)}=\sum_\nu e^{-(\beta_+E_+ + \beta_-E_-)}\,.
\eea
At this point, it is clear that taking the limit $\beta_-\rightarrow
\infty$ while $\beta_+\rightarrow \omega$ (constant), gives the
correct supersymmetric partition function. The above limiting
procedure takes $T$ to zero, {\it but} also scales $\Omega$ in such
a way that the new supersymmetric conjugated variable $\omega$ is
finite and arbitrary. Note that among all available states, only
those that satisfy the BPS bound are not suppress in the sum,
resulting in the supersymmetric partition function
\bea Z{(\omega)}=\sum_{bps} e^{-\omega J}\,, \eea
where the sum is over all supersymmetric states ($bps$) with $E=J$.
The above manipulations are easy to implement in more complicated
supersymmetric field theories like, {\it e.g.}, $N=4$ SYM theory in
four dimensions. What is less trivial is that amazingly it could
also be implemented in the dual supersymmetric configurations of
gauged supergravity, since it means that these extreme BPS solutions
are somehow protected from higher string theory corrections.

Before we apply the Euclidean zero-temperature formalism to concrete
black hole systems, it is profitable to highlight its key steps. To
study the statistical mechanics of supersymmetric black holes we
take the off-BPS BH solution and we send $T\rightarrow 0$. In this
limiting procedure, the angular velocities and electric potentials
at the horizon can be written as an expansion in powers of the
temperature. More concretely one has when $T\rightarrow 0$,
\bea  &&\beta \rightarrow \infty\,,\nn\\
 &&\Omega \rightarrow
\Omega_{bps}\,-\,{\omega\over\beta}\,+ \,O(\beta^{-2})\,,\nn\\
 && \Phi\rightarrow \Phi_{bps}\,-\,{\phi\over\beta}\,+ \,O(\beta^{-2})\,,
\label{multi}\eea
where $\beta$ is the inverse temperature; $(\Omega, \Phi)$ are the
angular velocities and electric potentials at the horizon; the
subscript $bps$ stands for the values of these quantities in the
on-shell BPS solution; and $(\omega,\phi)$ are what we call the
supersymmetric conjugated potentials, {\it i.e.}, the next to
leading order terms in the expansion. For all the systems studied,
we find that the charges have the off-BPS expansion,
\bea \label{gen1}  E=E^{bps}+\mathcal{O}\left( \beta^{-2} \right)
\,, \qquad Q = Q^{bps}+\mathcal{O}\left( \beta^{-2} \right)
\,,\qquad J = J_{\phi}^{bps} +\mathcal{O}\left( \beta^{-2}\right),
\eea
where $(E,Q,J)$ are the energy, charges and angular momenta of the
BH. In supergravity, the grand canonical partition function in the
saddle point approximation is related to so called quantum
statistical relation (QSR) \cite{Gibbons:2004ai}
\bea I_{(\beta,\Phi,\Omega)}=\beta E-\Phi Q-\Omega J-S\,,
\label{QSR:intro}\eea
where $S$ is the entropy, and $(\beta,\Phi,\Omega)$ are interpreted
as conjugated potentials to $E,Q,J$, respectively. $I$ is the
Euclidean action (evaluated on the off-BPS BH solution) that, in
this ensemble, depends only on $(\beta,\Phi,\Omega)$. It plays the
role of free energy divided by the temperature. Inserting
(\ref{multi}) and (\ref{gen1}) into (\ref{QSR:intro}) yields
\bea I_{(\beta,\Phi,\Omega)}=\beta (E^{bps}-\Phi_{bps}
Q^{bps}-\Omega_{bps} J^{bps})+\phi Q^{bps}+\omega J^{bps}-S_{bps}  +
\mathcal{O}\left( \beta^{-1}\right)\,. \label{QSR2:intro}\eea
Here, we observe that this action is still being evaluated {\it
off}-BPS. Moreover, the term multiplying $\beta$ boils down to the
BPS relation between the charges of the system and thus vanishes
(this will become explicitly clear in the several examples we will
consider). This is an important feature, since now we can finally
take the  $\beta\rightarrow \infty$ limit yielding relation
(\ref{sqsr}). With the present notation it reads as
\begin{equation}
I_{bps}=\phi Q^{bps}+\omega J^{bps}-S_{bps}\,.
\label{sqsr:intro}\end{equation}
It is important to stress that this zero temperature limiting
procedure yields a finite, not diverging, supersymmetric version of
QSR, or shortly SQSR. Note that if we had evaluated the Euclidean
action (\ref{QSR:intro}) directly on-shell it would not be well
defined, as is well-known. As a concrete realization, we picked (and
will do so along the paper) the SQSR to exemplify that the
$T\rightarrow 0$ limit yields well-behaved supersymmetric relations.
The reason being that this SQSR relation is the one that will
provide direct contact with Sen's entropy functional formalism,
which is the main aim of our study. However, it also provides a
suitable framework that extends to the study of the {\it full}
statistical mechanics of supersymmetric black holes.

\vspace{.5cm} \noindent {\bf Euclidean action and chemical
potentials of BPS D1-D5-P black holes}\vspace{.5cm}

As we pointed out in the introduction, due to the attractor
mechanism, BH in ungauged supergravity have a dual CFT theory
defined in the boundary of its near horizon geometry. Therefore, and
in a similar way as for asymptotic AdS spacetimes, these BH should
be related to statistical ensembles in the dual CFT. As a direct
consequence of this duality, we conclude that in the ungauged case
there should also exist a well defined zero temperature limit in the
supergravity description that yields the dual CFT chemical
potentials.

In what follows, we apply the Euclidean $T\rightarrow 0$ limit to
the illuminating example of five-dimensional three charged BH with
two angular momenta that can be described as the D1-D5-P system of
type II$B$ supergravity \footnote{We present this case as a main
example, but include many others in the Appendix
\ref{sec:several}.}. This solution can also be embedded as a
solution of eleven-dimensional supergravity, or as a solution of
type II$A$, where all these different descriptions are related by
dimensional reduction and $U$-dualities. A detailed review of the
D1-D5-P BH solution \cite{Cvetic:1996xz,Giusto:2004id} and its
thermodynamic properties needed for our discussion can be found in
Appendix \ref{sec:d1d5p}.

In type II$B$, the ten-dimensional system can be compactified to
five dimensions on $T^4 \times S^1$ with the D5-branes wrapping the
full internal space and the D1-branes and KK-momentum on the
distinguished $S^1$. The length of $S^1$ is $2\pi R$ and the volume
of $T^4$ is $V$. We will work in units such that the
five-dimensional Newton constant is $G_5=G_{10}/2\pi RV=\pi/4$. The
ten-dimensional solution is characterized by six parameters: a mass
parameter, $M$; spin parameters in two orthogonal planes,
$(a_1,a_2)$; and three boost parameters,
$(\delta_1,\delta_5,\delta_p)$, which fix the D1-brane, D5-brane and
KK-momentum charges. The physical range of $M$ is $M \geq 0$. We
assume without loss of generality that $\delta_i \geq 0$
($i=1,5,p$), and $a_1 \geq a_2 \geq 0$ (The solutions with
$a_1a_2\leq 0$ are equivalent to the $a_1a_2\geq 0$ ones due to the
symmetries of the solution). We will use the notation $c_i \equiv
\cosh \delta_i$, $s_i \equiv \sinh \delta_i$.

The BH charges  are: ADM mass $E$, the angular momenta
$(J_{\phi},J_{\psi})$ and the gauge charges $(Q_1, Q_5, Q_p)$
associated with the D1-branes, D5-branes and KK momentum. In terms
of the parameters describing the solution they are given by
\bea E &=& \frac{M}{2} \left[ \cosh (2\delta_1)
+ \cosh (2\delta_5) + \cosh (2 \delta_p)\right], \nonumber \\
J_{\phi} &=& -  M (a_2 c_1 c_5 c_p - a_1 s_1 s_5 s_p)\,,\nonumber \\
J_{\psi} &=& -  M (a_1 c_1 c_5 c_p - a_2 s_1 s_5 s_p)\,,\nonumber \\
 Q_i &=& M s_i c_i\,, \qquad i=1,5,p \,. \label{ADMcharges}
\eea
Note that these quantities are invariant under interchange of the
$\delta_i$'s. This reflects the equivalence of the several
geometries obtained by $U$-dualities, that also interchange the
several gauge charges.

Regarding the thermodynamical properties of these BH, it is
convenient for future use to define the left and right temperatures,
$T_L$ and $T_R$, through the relation
$\beta=\frac{1}{2}(\beta_L+\beta_R)$ ($\beta=1/T$ and
$\beta_{L,R}=1/T_{L,R}$). Then, using this relation together with
(\ref{roots}) on (\ref{beta}) yields\footnote{Expressions
(\ref{betaLR})-(\ref{electricpotent(M)}) agree with the ones first
computed in \cite{Cvetic:1998xh} upon the notation identification
$a_1\rightarrow -l_2$ and $a_2\rightarrow -l_1$.}
\bea \beta_L = \frac{ 2\pi M \left(c_1c_5c_p - s_1s_5s_p\right)}
                 {\left[ M-(a_2-a_1)^2 \right]^{1/2} }\,,\quad\quad
\beta_R = \frac{ 2\pi M \left(c_1c_5c_p + s_1s_5s_p\right)}
                 {\left[ M-(a_2+a_1)^2 \right]^{1/2} } \,. \label{betaLR}
\eea
The BH angular velocities $\Omega^{\phi,\psi}$ and electric
potentials $\Phi^{(i)}$ are computed in Appendix \ref{sec:d1d5p}.
Here, using (\ref{roots}), we rewrite them in terms of the
parameters $(M,\delta_1,\delta_5,\delta_p,a_1,a_2)$
\bea &&\hspace{-1cm}\Omega^{\phi,\psi} = -\frac{\pi}{\beta} \left[
 \pm \frac{a_2-a_1}{\left[M-(a_2-a_1)^2 \right]^{1/2}}
  +\frac{a_2+a_1}{\left[M-(a_2+a_1)^2 \right]^{1/2}} \right],
                                         \\  \label{velocity(M)}
&&\hspace{-1cm}\Phi^{(i)} = \frac{\pi M} {\beta} \left[
\frac{(\tanh\delta_i)c_1c_5c_p - (\coth\delta_i)s_1s_5s_p}
   {\left[M-(a_2-a_1)^2 \right]^{1/2}}
+ \frac{(\tanh\delta_i)c_1c_5c_p + (\coth\delta_i)s_1s_5s_p}
{\left[M-(a_2+a_1)^2 \right]^{1/2}} \right],
   \label{electricpotent(M)}
\eea
while the expression for the entropy is
\bea S = \pi M \left[  \frac{c_1c_5c_p + s_1s_5s_p}{\left[
M-(a_2-a_1)^2 \right]^{-1/2} } +\frac{c_1c_5c_p - s_1s_5s_p}{\left[
M-(a_2+a_1)^2 \right]^{-1/2}} \right] \,. \label{entropy(M)} \eea
The BPS limit of the three charged BH is obtained by taking
$M\rightarrow 0$, $\delta_i\rightarrow \infty$,
$J_{\phi}+J_{\psi}\rightarrow 0$ while keeping $Q_i$ fixed. In this
supersymmetric regime, the charges satisfy the BPS constraints
\bea E^{bps}= Q_1^{bps}+Q_5^{bps}+Q_p^{bps} \,,  \qquad
J_{\psi}^{bps} = -J_{\phi}^{bps}  \,. \label{ADMchargesBPS} \eea
As a first step to define the Euclidean $T\rightarrow 0$ limit, we
consider the near-BPS limit of this solution,
\bea J_{\phi}+J_{\psi}\rightarrow 0 \,;  \quad M\rightarrow 0 \,,
\quad \delta_{1,5} \rightarrow \infty \,,  \quad Q_{1,5} \:\:{\rm
fixed}\,;  \quad
 \delta_p \:\:{\rm finite} \,. \label{nearBPSlim}
\eea
That is, in the near-BPS limit we keep $\delta_p$ large {\it but}
finite. This limit is also often called the dilute gas regime since
we are neglecting the interactions between left and right movers.
Note that since the three charges can be interchanged by
$U$-dualities, it does not matter which one of the boosts we keep
finite. Given this equivalence we choose to keep $\delta_p$ finite,
without any loss of generality.

Now, to take the  $T\rightarrow 0$ limit, we define the off-BPS
parameter $\varepsilon$, that measures energy above extremality, to
be such that $E\equiv E^{\rm bps}+\varepsilon$. In terms of the
solution parameters it is given by $\varepsilon=M e^{-2\delta_p}/4$.
The details of the off-BPS expansion that we carry on in the sequel
can be found in Appendix \ref{sec:nearBPS}. Here we just quote the
relevant results. We can expand the left and right temperatures in
terms of the off-BPS parameter  $\varepsilon$ yielding,
\bea \beta_L =  \frac{\pi Q^{bps}_1 Q^{bps}_5}{\sqrt{ Q^{bps}_1
Q^{bps}_5 Q^{bps}_p -(J^{bps}_{\phi})^2}} \,,\qquad \beta_R = \pi
\sqrt{Q^{bps}_1 Q^{bps}_5}\frac{1}{ \sqrt{\varepsilon} } \,.
\label{expand:betaLR} \eea
So the BPS limit corresponds to send the temperature $T\rightarrow
0$ by sending $\beta_R \rightarrow \infty$ while keeping $\beta_L$
finite (we are left with only left-movers). Hence, we find more
appropriate to use $\beta_R$ as the off-BPS parameter instead of
$\varepsilon$ . These two quantities are related by the second
relation of (\ref{expand:betaLR}).

We can now expand all the thermodynamic quantities in terms of this
off-BPS quantity $\beta_R^{-1}$. For the angular velocities and
electric potentials, the expansion yields
\bea
 \Omega^{\phi,\psi} &=&\Omega_{bps}^{\phi,\psi}
 - \frac{2}{\beta_{R} }\frac{\mp \pi J_{\phi}^{bps}}
 { \left[ Q^{bps}_1 Q^{bps}_5 Q^{bps}_p -(J^{bps}_{\phi})^2 \right]^{1/2} }
  + \mathcal{O}\left( \beta_{R}^{-2}\right) \,, \nonumber \\
\Phi^{(i)} &=& \Phi_{bps}^{(i)}-\frac{2}{\beta_R}\frac{\pi
Q_1^{bps}Q_5^{bps} Q_p^{bps}} {Q_i^{bps}\left[ Q^{bps}_1 Q^{bps}_5
Q^{bps}_p -(J^{bps}_{\phi})^2 \right]^{1/2}}\ +\mathcal{O}\left(
\beta_R^{-2}\right)\,.
   \label{expand:potent}
\eea
where the BPS angular velocities and electric potentials are
\bea
 \Omega_{bps}^{\phi,\psi} &=& 0 \,; \qquad \Phi_{bps}^{(i)}=1\,.
   \label{potentialBPS}
\eea
The expansion of the conserved charges yields
\bea
&&E= E^{bps}  
  +\mathcal{O}\left( \beta_R^{-2} \right) \,, \qquad
J_{\phi} = J_{\phi}^{bps}
 +\mathcal{O}\left( \beta_R^{-2}\right)       \,,\qquad
J_{\psi} = -J_{\phi}^{bps}
 +\mathcal{O}\left( \beta_R^{-2}\right)
    \,,\nonumber \\
&& Q_1 \simeq Q_1^{bps}\,, \qquad Q_5 \simeq Q_5^{bps}\,, \qquad
 Q_p = Q_p^{bps}
 +\mathcal{O}\left( \beta_R^{-2}\right)\,. \label{expand:ADMcharges}
\eea
Note that the BPS charges satisfy (\ref{ADMchargesBPS}). They are
written in terms of the parameters that describe the system in
(\ref{ADMmovers}). Finally, the expansion of the entropy yields
\bea
S = S^{bps} 
  +\mathcal{O}\left( \beta_R^{-1} \right) \,, \qquad {\rm with}\qquad
S^{bps}=2\pi \left[ Q^{bps}_1 Q^{bps}_5 Q^{bps}_p
-(J^{bps}_{\phi})^2 \right]^{1/2}. \label{Sbps} \eea

With the above off-BPS expansion, we are ready to define the BPS
chemical potentials. Comparing (\ref{expand:potent}) with
(\ref{multi}) we obtain,
\bea \omega_{\phi,\psi}= \mp \frac{\pi J_{\phi}^{bps}} {
\left[Q^{bps}_1 Q^{bps}_5 Q^{bps}_p -(J^{bps}_{\phi})^2
\right]^{1/2} } \,,\quad\quad \phi_{i} = \frac{\pi
Q_1^{bps}Q_5^{bps} Q_p^{bps}} {Q_i^{bps}\left[ Q^{bps}_1 Q^{bps}_5
Q^{bps}_p -(J^{bps}_{\phi})^2 \right]^{1/2}}\,. \nn
\\\label{ConjPotential}
\eea
Notice that these chemical potentials only depend on the BPS
conserved charges.

Now that all the BPS statistical mechanics conjugated pairs and
entropy are defined, we are ready to obtain the other thermodynamic
functions. For example, consider the quantum statistical relation,
\bea I=\beta E -\beta\sum_{i=1,5,p} \Phi^{(i)} Q_i -\beta
\sum_{j=\phi,\psi} \Omega^{j}J_{j} -S\,.
    \label{QSR:d1d5p}
\eea
After the  off-BPS expansion, {\it i.e.}, using
(\ref{expand:ADMcharges}), (\ref{Sbps}) and (\ref{expand:potent}) it
yields
\bea &&I=\beta\left( E^{bps}-\sum_{i=1,5,p} Q_i^{bps}
-\sum_{j=\phi,\psi} \Omega_{bps}^{j}J_{j}^{bps} \right) +
\sum_{i=1,5,p} \phi_{i}\, Q_i^{bps} +\sum_{j=\phi,\psi}
\omega_{j}\,J_{j}^{bps}-S_{bps}
 +\mathcal{O}\left( \beta_R^{-1}\right)\,. \nonumber \\
 &&
    \label{QSR:aux}
\eea
The term in between brackets vanishes due to the BPS relations
(\ref{ADMchargesBPS}) and (\ref{potentialBPS}). Then, taking
$\beta\rightarrow \infty$, we are left with the supersymmetric
quantum statistical relation (SQSR) for the three-charged BH,
\bea I_{bps}=\phi_1\, Q_1^{bps}+\phi_5\, Q_5^{bps}+\phi_p\,
Q_p^{bps} +2\omega_{\phi} J_{\phi}^{bps} -S_{bps}\,,
 \label{SQSR:d1d5p}
\eea
where $I_{bps}$ is the value of the Euclidean action in the
supersymmetric limit of the D1-D5-P BH, and we  used $J^{bps}_{\psi}
=-J^{bps}_{\phi} $ and $\omega_{\psi}=-\omega_{\phi}$. Notice that
$I_{bps}$ corresponds to the Legendre transformation of the entropy
with respect to all the BPS chemical potential and therefore should
be interpreted as the BH free energy.

The off-BPS expansion of the horizon angular velocities and electric
potentials gives the supergravity chemical potentials as the next to
leading order term of the expansion around the BPS solution. The
motivation for this expansion analysis comes from the fact that BPS
BHs can be studied as dual configurations to supersymmetric
ensembles at zero temperature {\it but} non-zero chemical potentials
in the dual CFT \cite{SQSR1}. The supergravity conjugated potentials
(\ref{ConjPotential}) are then the strong coupling dual objects to
the CFT chemical potentials. The SQSR relation (\ref{SQSR:d1d5p})
will be connected to the well-known Sen's entropy relation in the
next section.

\setcounter{equation}{0}\section{Euclidean zero-temperature and
entropy functional formalisms} \label{sec:Map}

In previous sections we have described two apparently unrelated
procedures to obtain the entropy of supersymmetric BH that naturally
contain the definitions of pairs of conjugated variables, related to
the BH charges. In this section we show that both procedures produce
basically the same body of final definitions, even though
conceptually both approaches are rather different.

That both approaches produce the same final chemical potentials and
definitions can be seen in any of the examples at hand. As usual,
the best way to illustrate our point is to pick a system that
captures the fundamental ingredients, while avoiding features that
do not play a key role and produce unnecessary distraction from the
main point. In the present case, the appropriate system is the
non-rotating D1-D5-P BH (later, we will discuss the rotating case).
Comparing the thermodynamic relations (\ref{Sbps}),
(\ref{ConjPotential}), and the Sen's relations (\ref{sen}),
(\ref{Ssen}), we can indeed confirm that all the key quantities
agree in the two formalisms. Explicitly we have that
\bea \phi_i=2\pi e_i\,,\quad\quad\quad Q_i=q_i\,,\quad\quad\quad
I_{bps}=2\pi f\,. \label{phi:e} \eea

Nevertheless, that both frameworks are equivalent is {\it a priori}
not at all obvious since they have important differences. Sen's
approach relies completely on the structure of the near horizon
geometry. In particular, the entropy is constructed analyzing Wald's
prescription and Einstein equations in these spacetimes and all the
analysis is carried on at the BPS bound {\it i.e.}, when the
solution is extremal. In contrast, the zero temperature limit
approach relies on the thermodynamical properties of BH and, in
principle, uses the whole spacetime, not only the near horizon
region. The resulting thermodynamic definitions come as a limiting
behavior of non-extremal BH and have a nice straightforward
interpretation in terms of the dual CFT thermodynamics.

\subsection{Near-horizon and
asymptotic contributions to the Euclidean action} \label{sec:Map1}

To understand why the above close relations between the two
formalisms hold, let us go back to the calculation of the Euclidean
action for general BH in the {\it off}-BPS regime. Inspired in ten
dimensional type II supergravity, we start with the general
action\footnote{For simplicity, the reasoning is done at the level
of two derivative Lagrangian. Nevertheless, following Wald's
approach for higher derivative actions, we notice that the BH action
can always be recast as surface integrals. Moreover, for
definiteness, we anchor our discussion to type II action, but
whenever needed we make comments to extend our arguments to more
general theories.}
\bea I={1\over 16\pi G}
\int_\Sigma{\sqrt{-g}\left(R-{1\over2}(\partial\Psi)^2-{1\over 2
n!}e^{\alpha\Psi}F_{(n)}^2\right)}+{1\over 8\pi
G}\int_{\partial\Sigma}{K}\,,\eea
where $\Sigma$ is the spacetime manifold, $\partial\Sigma$ the
boundary of that manifold and $K$ is the extrinsic curvature. In the
BH case, once we have switched to Euclidean regime, it is necessary
to compactify the time direction to avoid a conical singularity.
This compactification defines the Hawking temperature as the inverse
of the corresponding compactification radius.

To evaluate the Euclidean action on the BH solution, one of the
methods to obtain a finite result, {\it i.e.}, to regularize and
renormalize the action, consists of putting the BH in a box and
subtract the action of a background vacuum solution
$(g^0,\Psi^0,F^0)$. This procedures also defines the ``zero" of all
the conserved charges. For asymptotic flat solutions we use
Minkowski, while for asymptotic AdS solutions we use AdS. Once in
the box, the radial coordinate is restricted to the interval
$(r_+,r_b)$, where $r_+$ is the position of the horizon and $r_b$
corresponds to an arbitrary point which limits the box and that at
the end is sent to infinity. Another important ingredient is the
boundary conditions on the box. Basically, depending on which
conditions we impose on the different fields, we will have fixed
charges or fixed potentials. If we do not add any boundary term to
the above action, we will be working with fixed potentials, {\it
i.e.}, we will work in the grand canonical ensemble \cite{Braden}.

The field equations are derived from a variational principle, where
fields are kept constant at the boundaries. In particular, the trace
the of equation that comes from the variation of the metric (for the
D1-D5-P system, see equation (\ref{FieldEqs})) implies that
\bea R-{1\over2}(\partial\Psi)^2 = a e^{\alpha\Psi}F_{(n)}^2\,, \eea
where $a$ depends on the spacetime dimensions and $n$. Therefore,
on-shell, the action reduces to\footnote{Where we have used that the
action of the background vacuum solution over $\Sigma$ is zero.},
\bea I={b\over 8\pi G}\int_\Sigma{e^{\alpha\Psi}F_{(n)}^2} + {1\over
8\pi G}\int_{\partial\Sigma}{\left(K-K^0\right)}\,,\eea
where $b$ depends on the spacetime dimensions and $n$. The first
term is a volume integral over $\Sigma$ that can easily be converted
into a boundary integral over $\partial\Sigma$, once we recall that
we are considering electric fields only and hence
$F_{(n)}=dC_{(n-1)}$. Integrating by parts we get
\bea I={c\over 8\pi
G}\int_{\partial\Sigma}{e^{a\Psi}F_{(n)}C_{(n-1)}} + {1\over 8\pi
G}\int_{\partial\Sigma}{\left(K-K^0\right)}\,, \label{euclidean}\eea
where $c$ depends on the spacetime dimensions and $n$. At this
point, the on-shell Euclidean action is completely recasted in two
surface integrals terms, evaluated at $r_+$ and $r_b$. Consider
first the extrinsic curvature term. At $r_b$, we get $\beta E_b$,
where $E_b$ is the quasi-local energy. When $r_b$ is taken to
infinity, $E_b$ reduces to the BH energy $E$ and we recover usual
term $\beta E$. At $r_+$, only $K$ contributes and gives minus the
area of the horizon divided by $4G$, {\it i.e.}, minus the
Bekenstein-Hawking entropy $S$. Next consider the first term. Here
the integral over time gives the factor $\beta$, while the
integration over the other directions (of the induced metric
determinant at the boundary times $e^{a\Psi}F_{(n)}$) gives the
corresponding electric charge $Q$. Therefore, we get
\bea {c\over 8\pi G}\int_{\partial\Sigma}{e^{a\Psi}F_{(n)}C_{n-1}}=
-\beta Q\left[C_{(n-1)}(r_b)-C_{(n-1)}(r_+)\right]\,. \eea
Then, we use the definition of the conjugated chemical potential
$\phi$ as the difference of the gauge potential at infinity and at
the horizon,
\bea \Phi=C|_\infty-C|_{r_+}\,, \eea
and hence, when $r_b$ is sent to infinity, we recover the usual term
$-\beta Q \Phi$. As a grand total we obtain the QSR,
\bea I=\beta E-\beta \Phi Q -S\,.\eea
Now, it is important to notice that the definition of $\Phi$ is
gauge independent, and therefore we can always choose a particular
gauge that simplifies the picture depending on which physical
concepts we want to stress. Here,  we choose the ``natural gauge"
adapted to the BPS limiting cases, $C|_\infty=\Phi_{bps}$, where
$\Phi_{bps}$ is usually $1$ in natural units and for asymptotically
flat BHs. Note that in this gauge one has
$C|_{r_+}=\Phi_{bps}-\Phi$. This gauge choice is the one that makes
direct contact between the Euclidean zero temperature and entropy
function formalisms for reasons that will become clear after
(\ref{split2}).

At this point we are ready to rewrite the Euclidean action in two
pieces, one evaluated in the first boundary at $r=r_+$, and the
other in the second boundary at $r=\infty$,
\bea I=\int_{r=r_+}\left\{ {c\over 8\pi G}e^{a\Psi}F_{(n)}C_{(n-1)}
+ {1\over 8\pi G}K\right\} + \int_{r=\infty}\left\{ {c\over 8\pi
G}e^{a\Psi}F_{(n)}C_{(n-1)} + {1\over 8\pi
G}\left(K-K^0\right)\right\} \,.  \eea
Evaluating both terms as we did before but now in the adapted gauge
we get,
\bea I=\underbrace{\beta(\Phi_{bps}-\Phi)Q-S} \,+\,
\underbrace{\beta(E-\Phi_{bps}Q)}\,.\\
r=r_+\quad\quad\quad\quad\quad\: r=\infty\quad\quad \nn
\label{actionsplit}\eea
Therefore we can always find a gauge in which the Euclidean action
splits in two contributions, one at the horizon and the other in the
asymptotic region. It is perfectly adapted to understand the near
horizon regime. Equally interesting, this expression is also adapted
to understand the supersymmetric limit. In fact, from our discussion
in section \ref{sec:QSRflat}, it is easy to see that the first term
exactly reproduces the SQSR, {\it i.e.},
\bea \lim_{BPS\;limit}\; \beta(\Phi_{bps}-\Phi)Q-S = \phi
Q_{bps}-S_{bps}\,. \label{split1}\eea
On the other hand, the asymptotic term vanishes due to fact that
$\Phi_{bps}=1$, and thus the leading term in the expansion is
nothing else than the BPS relation $E_{bps}=Q_{bps}$ characteristic
of supersymmetric regimes, {\it i.e.},\footnote{This discussion is
strictly valid for the asymptotically flat BHs where $\Phi_{bps}=1$.
For asymptotically AdS BHs, the normalization usually chosen in the
literature yields in general $\Phi_{bps}\neq 1$. However, in this
case, the term inside brackets in (\ref{split2}) still vanishes
because it is exactly the BPS constraint on the charges. This
follows by construction and is explicitly confirmed for 5$D$ gauged
supergravity in \cite{SQSR1,Silva:2007tw}.}
\bea \lim_{BPS\;limit}\; \beta(E-\Phi_{bps}Q)=\lim_{BPS\;limit}\;
\beta(E-Q) = 0 \label{split2}\,.\eea
(Note that in the last equality we jump some steps that were already
explained in detail after (\ref{QSR2:intro}), and that we do not
repeat here. They guarantee that this term indeed vanishes and does
not give an indeterminacy of the type $\infty\cdot 0$). {\it We
conclude that the Euclidean action of the BH at the BPS bound is
given exclusively from the near horizon part of the solution}. This
is another way to characterize the attractor mechanism, since the
physical properties of the solution are captured entirely by the
near horizon geometry. From the above result, it is easy to see why,
for supersymmetric cases, $I$ is related to $f$. First, both are
functionals of the near horizon geometry alone. Also, the time and
radial integrations are trivial and only integration on the other
space directions actually contribute. In fact, this is a way to
understand why in the definition of $f$ there is no integration in
the AdS part of the near horizon metric. Note also that in Sen's
approach the $f$ function is defined as the integral of the {\it
string frame} Lagrangian evaluated at the near horizon geometry.
Since in this geometry the dilaton is a constant, the string frame
and Einstein frame Lagrangians are related by a trivial constant
factor.

We now discuss the effects introduced by addition of rotation.
Working in a coordinate system in which the geometry is not rotating
at infinity, the action can be splited as
\bea
I=\underbrace{\beta(\Phi_{bps}-\Phi)Q+\beta(\Omega_{bps}-\Omega)J-S}
\,+\,
\underbrace{\beta(E-\Phi_{bps}Q-\Omega_{bps}J)}\,.\\
r=r_+\qquad\qquad\qquad\qquad\quad\quad\:\:\: r=\infty\qquad\qquad
\! \nn \label{Rot:actionsplit}\eea
By definition, the near horizon term contains  all the information
on the chemical potentials (once the BPS limit is taken),
\bea \lim_{BPS\;limit}\;
\beta(\Phi_{bps}-\Phi)Q+\beta(\Omega_{bps}-\Omega)J-S = \phi
Q_{bps}+\omega J_{bps}-S_{bps}\,, \label{Rot:split1}\eea
while the asymptotic term is again the BPS constraint between the
several charges and thus vanishes,
\bea \lim_{BPS\;limit}\; \beta(E-\Phi_{bps}Q-\Omega_{bps}J)=0\,.
\label{Rot:split2}\eea
For asymptotically flat BHs one always has $\Omega_{bps}=0$ and
(\ref{Rot:split2}) reduces to (\ref{split2}). The horizon of flat
BHs does not rotate (angular momentum comes from the Poynting vector
of electromagnetic fields) and this is one way to understand why the
angular momenta does not appear in their BPS constraint. On the
other hand, the horizon velocity of asymptotically AdS BHs is, in
general, non-vanishing, and thus the angular momenta also
contributes to the BPS constraint of these BHs.

\subsection{Relation between chemical potentials in the two
formalisms} \label{sec:Map2}

At this point only reminds to understand the relation between the
conjugated potentials in both pictures. In Sen's approach, the
information about them is contained in the electric fields of the
near horizon geometry, while in the Euclidean zero temperature
formalism this information is encoded in the next to leading order
term in an off-BPS expansion of the full geometry. Although these
definitions seem to be rather different at first sight, notice that
in Sen's approach the field strength is just the radial derivative
of the potential evaluated at the horizon. In the Euclidean zero
temperature case, the off-BPS expansion can be rewritten as an
expansion in the radial position of the horizon $\rho_+$. Therefore,
the next to leading order term in the off-BPS expansion of the gauge
potential at $\rho_+$ is proportional to its derivative with respect
to the radial position of the horizon. Hence it is proportional to
the field strength at the horizon. These words can be made very
precise by taking an example. Consider the D1-D5-P BH we have been
working with (again we do not include rotation in the analysis to
avoid unnecessary non-insightful complications). In the full
geometry (\ref{norotmetric}), where the zero temperature limit
procedure is applied, we work with the $t,r$ coordinates. Sen's
approach uses instead the near-horizon fields (\ref{NHsen0}) or
(\ref{NHsen}) described in terms of $(\tau,\rho)$ coordinates. The
two set of coordinates are related by (\ref{coordtransf}). Our
purpose in the next lines is to understand the first relation in
(\ref{phi:e}). For definiteness we focus on the relation
$\phi_1=2\pi e_1$. From (\ref{norotmetric}),
$C_{ty}=-Ms_1c_1/(\rho+Ms_1^2)$, and one also has the relation
between the gauge field written in the two coordinate systems,
$C_{ty}=\frac{\partial\tau}{\partial t}\,C_{\tau y}$. In the
near-horizon approach, the expression for $e_1$ comes from the
radial derivative of the potential evaluated at the BPS horizon
($\rho_+^{bps}=0$):
\bea e_1= {\biggl \{} \partial_{\rho} C_{\tau y}
 {\biggr \}}_{\rho=\rho_+^{bps}}={1\over 2}\sqrt{Q_5^{bps}Q_p^{bps}\over Q_1^{bps}}\,.
 \label{e:eq1} \eea
In the Euclidean zero temperature approach, the electric potential
is obtained by contracting the gauge field with the timelike Killing
vector $\xi=\partial_t$ yielding:
 $\Phi^{(1)}=-C_{ty}{\bigr |}_{\rho=\rho_+} =-\frac{\partial\tau}{\partial t}\,C_{\tau
y}{\bigr |}_{\rho=\rho_+} $ (note that $\rho_+=\rho_+^{bps}=0$ only
in the BPS case). As is clear from (\ref{expand:potent}), our
conjugated potential is defined as
\bea \phi_1= -\frac{1}{2}{\biggl \{} \frac{\partial
\Phi^{(1)}}{\partial \beta_R^{-1}}
 {\biggr \}}_{\beta_R^{-1}=0} =
 -\frac{1}{2} \frac{\partial \rho_+}{\partial \beta_R^{-1}}{\biggl \{}
\frac{\partial \Phi^{(1)}}{\partial \rho_+}
 {\biggr \}}_{\rho_+=\rho_+^{bps}} \label{e:eq2}  \,. \eea
Note the following key relations,\footnote{The presence of the
factor $1/2$ in the last equality is due to a subtlety that occurs
when we take $\partial_{\rho_+}\Phi$ (and thus before sending
$\rho_+ \rightarrow \rho_+^{bps}$). In the large $\delta_1$ regime
one has $Ms_1^2\sim Q_1^{bps}-M/2$. Using this and $\rho_+=M$
yields, in the denominator of $\Phi$, $\rho_+ + Ms_1^2\sim
Q_1^{bps}+\rho_+/2$. This is the $1/2$ that appears when we further
take the $\rho_+$ derivative. Note that this factor does not appear
in the last derivative of (\ref{e:eq3}), $\partial_{\rho} C_{\tau
y}$, because here we take the radial derivative evaluated on the
on-shell solution $\rho=0$.}
\bea {\biggl \{} \frac{\partial \Phi^{(1)}}{\partial \rho_+}
 {\biggr \}}_{\rho_+=\rho_+^{bps}}
 =-\frac{\partial\tau}{\partial t}
 {\biggl \{} \partial_{\rho_+} \left[ C_{\tau
y}{\bigr |}_{\rho=\rho_+} \right] {\biggr \}}_{\rho_+=\rho_+^{bps}}
 = -\frac{1}{2} \frac{\partial\tau}{\partial t}
 {\biggl \{} \partial_{\rho} C_{\tau y}{\biggr \}}_{\rho=\rho_+^{bps}}
  \,. \label{e:eq3} \eea
From (\ref{e:eq1})-(\ref{e:eq3}), one finally has
\bea \phi_1=\frac{1}{4} \frac{\partial \rho_+}{\partial
\beta_R^{-1}} \frac{\partial\tau}{\partial t}\, e_1=2\pi e_1 \,.
\label{e:eq4} \eea
The last equality follows from (\ref{coordtransf}), and from
$\rho_+=M=4\pi \sqrt{Q_1^{bps}Q_5^{bps}Q_p^{bps}}\beta_R^{-1}$ (see
(\ref{expand:betaLR}) and the last statement of Appendix
\ref{sec:nearBPS}). Physically we can understand it by noting that
the near-horizon coordinates are precisely the ones appropriate to
find the value of the temperature, that avoids the standard conical
singularity in the Euclidean near-horizon geometry. An analysis
along the lines carried here for this specific case can be carried
on for general cases and yield the relations (\ref{phi:e}) between
the conjugated potentials found using the two formalisms.

\vspace{.5cm}
 To summarize, we have seen that for supersymmetric BH, the
Euclidean action and all the chemical potentials are defined in the
near horizon geometry. The asymptotic region would contribute only
in off-BPS cases. We have also shown why the chemical potentials are
proportional to the electric fields in the near horizon region, and
ultimately, we have understood, from the BH thermodynamics, the
emergence of Sen's entropy function as the extremal limit of the
quantum statistical relation or SQSR. As a bonus, we can now extend
the statistical mechanics analysis like the SQSR to BPS solutions
with no off-BPS known extension, because we have learned how to
calculate the relevant chemical potentials directly in the BPS
regime with no need of the limiting procedure.

\setcounter{equation}{0}\section{Extremal (non-BPS) black holes}
\label{sec:Extreme}
So far we have seen that two completely different procedures, namely
the Euclidean zero temperature formalism and Sen's entropy formalism
allow to compute the entropy and conjugated chemical potentials of
{\it supersymmetric} BHs. This is not an accident as proved in the
previous section. Now, as is well-known, Sen's approach also allows
to find the attractor values of {\it non}-BPS extreme BHs
\cite{Astefanesei:2006dd,Astefanesei:2006sy}. So a question that
naturally raises is if whether or not the Euclidean zero temperature
approach is also able to deal successfully with these type of
solutions. In this section we address this issue.

It is straightforward to conclude that the Euclidean formalism
indeed allows to find the chemical potentials of non-BPS extreme
configurations. This follows from an analysis similar to the
derivation presented in section \ref{sec:Map}, but this time
slightly modified to account for the fact that the extreme BH is not
BPS. Choosing the gauge $C|_\infty=\Phi_{ext}$ (and thus
$C|_{r_+}=\Phi_{ext}-\Phi$), the extreme analogue of
(\ref{Rot:actionsplit}) is
\bea
I=\underbrace{\beta(\Phi_{ext}-\Phi)Q+\beta(\Omega_{ext}-\Omega)J-S}
\,+\,
\underbrace{\beta(E-\Phi_{ext}Q-\Omega_{ext}J)}\,.\\
r=r_+\qquad\qquad\qquad\qquad\quad\quad\:\:\: r=\infty\qquad\qquad
\!
 \nn \label{Ext:actionsplit}\eea
where the first term boils down to the extreme counterpart of
(\ref{Rot:split1}),
\bea \lim_{ext.\;limit}\;
\beta(\Phi_{ext}-\Phi)Q+\beta(\Omega_{ext}-\Omega)J-S = \phi
Q_{ext}+\omega J_{ext}-S_{ext}\,, \label{Ext:split1}\eea
containing all the information on the chemical potentials.

On the other hand, for non-BPS extreme solutions, we find that the
asymptotic term in (5.1),
\bea \lim_{ext.\;limit}\; \beta(E-\Phi_{ext}Q-\Omega_{ext}J)\,,
\label{Ext:split2}\eea
in general does not vanish, as oppose to its BPS cousin. However, we
find the following important feature, at least in the cases we
studied: i) the cases where (\ref{Ext:split2}) does not vanish
correspond to extreme rotating solutions that have in common the
presence of an ergoregion; (ii) rotating extreme solutions without
ergosphere and non-rotating extreme solutions have vanishing
(\ref{Ext:split2}). This occurs at least on the three-charged,
four-charged and Kerr-Newman systems. In the cases where it vanishes
we again have that the Euclidean action of the BH at the extreme
bound is given exclusively from the near horizon part of the
solution. The physical properties of the solution are captured
entirely by the near horizon geometry, which makes the attractor
mechanism manifest \footnote{This discussion is at the level of two
derivative Lagrangian. If corrections are added, we expect that the
asymptotic part vanishes producing a finite result, also for extreme
BH with ergoregion.}.

In the above extremal non-BPS cases, we can explicitly verify that
the two formalisms indeed yield the same results. For this exercise
and as an example, we will discuss below two extreme three-charged
BH (whose BPS cousin was studied in the previous sections). To
emphasize that the relation between the Euclidean and Sen's
formalism is universal and not restricted to the three-charged
system, in Appendices \ref{sec:4Q} and \ref{sec:KN}, we further
extend the exercise to three other non-trivial extreme solutions
whose properties have been studied within Sen's formalism.

\subsection{Extreme three-charged black hole with
ergoregion} \label{sec:Extreme:ergo3Q}

In the D1-D5-P solution described by (\ref{3charge})-(\ref{rr}) we
can take, instead of the BPS limit described in section
\ref{sec:QSRflat}, a different limit that yields an extreme (but not
BPS) BH with an ergoregion. This is a case in which the system shows
only partial attractor mechanism.

Concretely, we take the near-extreme limit
\bea M\rightarrow (a_1+a_2)^2 +\varepsilon \,, \quad \varepsilon\ll
1 \,. \label{3Q:nearEXTlim:ergo} \eea
When the off-extreme parameter $\varepsilon$ vanishes, the
temperature  indeed vanishes since $\beta_R \rightarrow \infty$ in
(\ref{betaLR}). The off-extreme expansion of the conserved charges
(\ref{ADMcharges}) around the corresponding extreme values (obtained
by replacing $M$ by $(a_1+a_2)^2$ in (\ref{ADMcharges})) is
straightforward, and the expansion of the thermodynamic quantities
(\ref{betaLR})-(\ref{entropy(M)}) yields
\bea & & \beta_L =  \pi \left(c_1 c_5 c_p - s_1 s_5 s_p \right)
\frac{(a_1+a_2)^2}{\sqrt{a_1a_2}}
       +\mathcal{O}\left( \varepsilon \right)
 \,,\qquad
\beta_R =  2\pi (a_1+a_2)^2 \left(c_1 c_5 c_p + s_1 s_5 s_p \right) \frac{1}{ \sqrt{\varepsilon} } \,, \nonumber \\
& & S = S^{ext}
  +\mathcal{O}\left( \beta_R^{-1} \right) \,, \qquad
\Omega^{\phi,\psi}=
\Omega_{ext}^{\phi,\psi}-\frac{2\omega_{\phi,\psi}}{\beta_R}
+\mathcal{O}\left( \beta_R^{-2}\right) \,,
                                       \nonumber\\
& & \Phi^{(i)} = \Phi_{ext}^{(i)}-\frac{2 \phi_{i} }{\beta_R}
+\mathcal{O}\left( \beta_R^{-2}\right)\,, \qquad i=1,5,p \,,
   \label{Ergo:expand:3Q}
\eea
where the extreme values satisfy
\bea & & S_{ext}= 2\pi \sqrt{a_1a_2} (a_1+a_2)^2  \left(c_1 c_5 c_p
+ s_1 s_5 s_p \right)
                  \,, \nonumber \\
&& \Omega_{ext}^{\phi}=\Omega_{ext}^{\psi}= - \left[
(a_1+a_2)(c_1c_5c_p+s_1s_5s_p) \right]^{-1}
                                   \,, \nonumber \\
 & & \Phi_{ext}^{(i)}= \frac{
       (\tanh\delta_i)c_1c_5c_p + (\coth\delta_i)s_1s_5s_p}
     {c_1c_5c_p +s_1s_5s_p} \,,  \quad i=1,5,p        \,,
    \label{Ergo:extS:3Q}
\eea
and the conjugated potentials are
\bea
 & & \omega_{\phi,\psi}= -\frac{\pi}{2}
 \left[ \frac{a_1+a_2}{\sqrt{a_1a_2}}\frac{c_1c_5c_p-s_1s_5s_p}{c_1c_5c_p+s_1s_5s_p}
 \pm (a_1-a_2)\sqrt{a_1a_2}\right]
                                   \,,  \\
 & &  \phi_{i} = -\frac{\beta_L}{s_i c_i} \left[ \tanh\delta_1\tanh\delta_5\tanh\delta_p
 -\coth\delta_1\coth\delta_5\coth\delta_p \right]^{-1}  \,, \qquad
i=1,5,p \,. \nonumber
    \label{Ergo:conjPot:3Q}
\eea
These expressions for the potentials could be rewritten only in
terms of the conserved charges as expected by the attractor
mechanism. We avoid doing it because the expressions are long and
non-insightful.

The QSR for this system is
\bea  I&=& \beta\left( E^{ext}-\sum_{i=1,5,p}
\Phi_{ext}^{(i)}Q_i^{ext}
  -\Omega_{ext}^{\phi}J_{ext}^{\phi}-\Omega_{ext}^{\psi}J_{ext}^{\psi}\right)
  \nonumber \\
&& +\sum_{i=1,5,p}\phi_i\, Q_i^{ext}+
\omega_{\phi}J_{ext}^{\phi}+\omega_{\psi}J_{ext}^{\psi} -S_{ext}
+\mathcal{O}\left( \beta_R^{-1}\right)\,.
 \label{Ergo:QSR:3Q}
\eea
In the supersymmetric system the analogue of the term in between
brackets vanishes due to the BPS constraint on the conserved
charges. But, in general, for non-BPS extreme BHs it does not vanish
(see discussion associated with (\ref{Ext:split2})). In the present
case the factor in between brackets is $\frac{(a_1+a_2)^2}{2}
\frac{c_1c_5c_p-s_1s_5s_p}{c_1c_5c_p+s_1s_5s_p}$. Note that this
quantity vanishes when rotation is absent ($a_1=a_2=0$). When it is
present, the solution has an ergoregion and the non-vanishing
contribution is associated with its existence. Notice that in this
case the Euclidean action is not well-defined but, nevertheless, the
chemical potentials (\ref{Ergo:conjPot:3Q}) take finite values and
are physically relevant.

\subsection{Extreme three-charged black hole without
ergoregion } \label{sec:Extreme:Noergo3Q}

The metric of the D1-D5-P system is also a solution of type I
supergravity. A fundamental difference between  type II$B$ and type
I theories is that the later theory has half of the supersymmetries
of type II$B$. This feature implies that in type I, if we reverse
the sign of the momentum in the BPS D1-D5-P black hole, we get a
distinct solution that is extreme but non-BPS. We study this
solution of type I in this subsection, as the main example of an
extreme non-BPS solution without ergoregion where attractor
mechanism is fully manifest.

The near-extreme limit we now consider  is similar to the near-BPS
limit (\ref{nearBPSlim}) in which we send the boosts to infinity;
the difference being that now we take one of the boosts to be {\it
negative} (again, by $U$-dualities it does not matter which one).
The reason why these two limits are indeed different and, in
particular, why one of them yields a BPS BH and the other not is the
following \cite{Dabholkar:2006tb}. The three-charged BH describes,
in the supergravity approximation and after dualities, the F1-NS5-P
system that is a configuration of heterotic string theory
compactified on $T^4\times S^1$. We can describe this system as an
effective fundamental string with winding number $n_1 n_5$ (where
$n_1$, $n_5$ are the numbers of $F1$ and $NS5$ constituents), and
with momentum excitations traveling along it. Now, heterotic string
theory is chiral. Hence, the direction of the momentum along the
fundamental string sets if the solution is supersymmetric or not. In
our conventions, the supersymmetric configuration F1-NS5-P is the
one with no right-movers. So, in the supergravity approximation, the
BPS BH that describes this system is obtained by taking
$\delta_p\rightarrow +\infty$. But we can also consider the
heterotic string configuration with only right-movers. Due to the
chirality property, this F1-NS5-$\bar{\rm P}$ configuration is then
not supersymmetric. And the corresponding supergravity solution
obtained by taking $\delta_p\rightarrow -\infty$ is not a BPS BH.
Note that this solution is however extreme, {\it i.e.}, it has zero
temperature. The reason being that there are no left-movers to
collide with the right-movers and generate the closed string
emission that describes the Hawking radiation.

So we take the near-extreme limit ($\delta_{1,5}>0$; $\delta_p<0$,
$Q_p<0$) \footnote{ The rotation parameters in this limit go as
\bea a_{1,2} =- \sqrt{M} \frac{J_{\phi}^{\rm ext}}{2\sqrt{-Q^{ext}_1
Q^{ext}_5 Q^{ext}_p}} \left[ 1 
+\mathcal{O}\left( \varepsilon^{} \right) \right] \,.
\label{ext:expand:a1a2+} \eea
For comparison, in the BPS limit $a_{1,2}$ go instead as
(\ref{expand:a1a2+}).
 }:
\bea J_{\phi}-J_{\psi}\rightarrow 0 \,;\qquad M \rightarrow 0 \,,
\qquad \delta_{1,5} \rightarrow \infty \,, \quad Q_{1,5} \:\:{\rm
fixed}\,; \qquad
 \delta_p<0 \:\:{\rm finite}\,.
 \label{ext:nearBPSlim}
\eea
The conserved charges of the non-extreme three-charged BH are listed
in (\ref{ADMcharges}), and the temperature, entropy, and angular
velocities and potentials at the horizon are given in
(\ref{betaLR})-(\ref{electricpotent(M)}).

The charges in the extreme solution satisfy the constraint
\bea E^{ext}= Q_1^{ext}+Q_5^{ext}-Q_p^{ext}  \,,  \qquad
J_{\psi}^{ext} = J_{\phi}^{ext}  \,,
 \label{ext:ADMchargesBPS}
\eea
where we used $Q_p^{ext}=-M e^{-2\delta_p}/4$.

The off-extreme parameter, $\varepsilon=M e^{2\delta_p}/4$, measures
energy above extremality and is such that $E\equiv E^{\rm
bps}+\varepsilon$. The expansion of the left and right temperatures
in terms of the off-extreme parameter $\varepsilon$ yields,
\bea \beta_L = \pi \sqrt{Q^{ext}_1 Q^{ext}_5}\frac{1}{
\sqrt{\varepsilon} }\,, \qquad \beta_R = \pi Q^{ext}_1 Q^{ext}_5
\left[ -Q^{ext}_1 Q^{ext}_5 Q^{ext}_p -(J^{ext}_{\phi})^2
\right]^{-1/2}
 \,. \label{ext:expand:betaLR}
\eea
The extreme limit corresponds to send the temperature $T\rightarrow
0$ by sending $\beta_L \rightarrow \infty$ while keeping $\beta_R$
finite. In this limit there are no left-movers, only right-movers.
The first relation in (\ref{ext:expand:betaLR}) defines
$\varepsilon$ in terms of $\beta_L$.

The expansion for the relevant thermodynamic quantities is
\bea
& & S = S_{ext} 
  +\mathcal{O}\left( \beta_L^{-1} \right) \,, \qquad  \Omega^{\phi,\psi}=\Omega_{ext}^{\phi,\psi}-
 \frac{2\omega_{\phi,\psi}}{\beta_L}
                +\mathcal{O}\left( \beta_L^{-2}\right) \,,
 \nonumber \\
& & \Phi^{(i)} = \Phi_{ext}^{(i)}-\frac{2 \phi_{i} }{\beta_L}
+\mathcal{O}\left( \beta_L^{-2}\right)\,, \qquad i=1,5,p \,,
   \label{ext:expand:potent}
\eea
where
\bea & & S_{ext}= 2\pi \left[ -Q^{ext}_1 Q^{ext}_5
Q^{ext}_p-(J_{\phi}^{ext})^2 \right]^{1/2}
                  \,, \nonumber \\
& &  \Omega_{ext}^{\phi,\psi}=0\,, \qquad \Phi_{ext}^{(1,5)}= 1\,,
\qquad \Phi_{ext}^{(p)}= -1\,.
    \label{ext:ext}
\eea
The conjugated potentials are
\bea &&\omega_{\phi,\psi} = -\frac{\pi J^{ext}_{\phi}}
{\left[ -Q^{ext}_1 Q^{ext}_5 Q^{ext}_p -(J^{ext}_{\phi})^2 \right]^{1/2}}\,, \nonumber\\
&& \phi_{i} = -\frac{\pi  Q^{ext}_1 Q^{ext}_5 Q^{ext}_p }
  { Q_i^{bps}\left[ -Q^{ext}_1 Q^{ext}_5 Q^{ext}_p-(J^{ext}_{\phi})^2 \right]^{1/2} } \,, \qquad
i=1,5,p \,.
    \label{ext:ConjPotential}
\eea

Although this is a non-BPS solution, it satisfies the extremal
constraint (\ref{ext:ADMchargesBPS}) that is linear in the charges.
This, together with (\ref{ext:ext}), has the consequence that
(\ref{Ext:split2}) applied to this system vanishes, and the QSR for
this system simplifies to
\bea I_{ext}=\sum_{i=1,5,p} \phi_i\, Q_i^{ext}+2\omega_{\phi}
J^{ext}_{\phi} -S_{ext}\,,
 \label{ext:QSR}
\eea
where we used $J^{ext}_{\psi} =J^{ext}_{\phi} $ and
$\omega_{\psi}=\omega_{\phi}$.

So, contrarily to the example of the previous subsection, where the
system showed only partial attractor mechanism due to the existence
of an ergoregion, in the present system the ergoregion is absent and
the attractor mechanism is fully manifest, even though the extreme
solution is not BPS.

\setcounter{equation}{0}\section{Discussion} \label{sec:Discussion}

First of all, we would like to stress again the logic behind our
approach: zero temperature limits to reach extremal configurations
are naturally defined in statistical analysis of quantum field
theories. The AdS/CFT correspondence then requires that there has to
be a dual analysis for strings in AdS. Supergravity is just the tree
level part of the above theory, and thus we do not expect in general
a well defined zero temperature limit at this level. Here, by well
defined we mean a limit that generates a finite Euclidean action
when $T\rightarrow 0$. Nevertheless, we have found extremal BH that
seem to be protected, and therefore have a well defined zero
temperature limit. In some of these cases, the protection is based
on supersymmetric arguments but, in other cases, we just have
extremal non-BPS BH where in fact it is not well understood why
supergravity is giving the correct answer.

In this article we have applied the Euclidean zero temperature
formalism to supergravity solutions where Sen's formalism is well
understood. In doing so, we have shown that this method agrees with
Sen's entropy formalism, producing the same statistical mechanics
functions like the entropy and the chemical potentials. On the top
of this, the Euclidean zero temperature formalism has the key
advantage of connecting the entropy functional with the statistical
mechanics of the dual CFT and with the more canonical BH
thermodynamics.

More concretely, due to the attractor mechanism, we found that the
Euclidean action is itself given by the near horizon geometry alone,
and therefore can be connected to Sen's approach to calculate the
entropy. We showed how to relate all the different definitions in
both approaches and why they match. In particular, we are able to
understand the CFT dual of Sen's approach, using the established map
for the corresponding quantum statistical relation. For example,
Sen's function $f$ (evaluated on-shell) is nothing more than the BPS
limit of the Euclidean action and therefore is related to the dual
CFT partition function. The above relation is relevant for the OSV
conjecture \cite{oguri}, since now $I_{bps}$ or $f$ naturally takes
the place of free energy of the supersymmetric BH.

We also worked out the extension to extremal but non-supersymmetric
BH. Here, since we are dealing with two derivative Lagrangians, we
divide BH in two groups: those with ergoregion and those without it.
In all the cases with no ergoregion we have checked, the zero
temperature limit produces a well defined QSR at extremality, where
all the chemical potentials, entropy and the Euclidean action are
related to Sen's approach. This is not a triviality, since here
there is no supersymmetry to protect these tree-level results. This
resuly seems to imply a protection mechanism in the extremal case,
as suggested in \cite{Dabholkar:2006tb}.

In the other case of extremal BH with ergoregions, we found an
ill-defined limit, where the asymptotic contribution to the
Euclidean action diverges. Nevertheless, the near horizon
contribution is well behaved and produces the correct entropy and
chemical potentials. These results are in agrement with Sen's
approach since these geometries are not fully attracted. Therefore
they depend also on asymptotic values of the moduli. We interpret
this result as a confirmation that these geometries do receive
corrections from string theory that in turn will modify the
asymptotic region, and thus asymptotic charges like the energy. In
fact, in \cite{Emparan} rotating BH of this sort were studied
finding that for the ergoregion branch, the entropy, but not the
energy, could be matched with the microscopic CFT.

We would like to point out that although we worked with standard
low-energy supergravity, the inclusion of higher derivative terms
should not spoil the results. In the Euclidean approach, one now has
to compute the Euclidean action with the modified Lagrangian and
define the entropy as its Legendre transform with respect to the BPS
chemical potentials. This should give the same entropy as defined by
Wald (see \cite{comparison}).

The zero temperature limit analysis of supersymmetric CFT ensembles
motivated the corresponding analysis in the dual supergravity
system. In this paper our main goal was to make direct contact
between this formalism and Sen's entropy function approach. The
Euclidean zero temperature formalism further allows to scan the
phase structure of BH. A paradigm on the useful information that
this formalism allows to find about the CFT living on the boundary
of a BH geometry can be found in \cite{SQSR1,SQSR2}. It would be
interesting to make a similar application, this time to study the
CFT of the BH systems discussed in this paper.

\section*{Acknowledgments}

\noindent

 We acknowledge Roberto Emparan and Pau Figueras for a
critical reading of the manuscript.

\noindent This work was partially funded by the Ministerio de
Educacion y Ciencia under grant FPA2005-02211 and by Fundac\~ao para
a Ci\^encia e Tecnologia through project PTDC/FIS/64175/2006. OD
acknowledges financial support provided by the European Community
through the Intra-European Marie Curie contract MEIF-CT-2006-038924,
and CENTRA for hospitality while part of this work was done.




\appendix

\section*{Appendices}

\setcounter{equation}{0}

\setcounter{equation}{0}\section{\label{sec:d1d5p} Three-charged
black hole: solution and thermodynamics}

\subsection{\label{sec:d1d5pBH} The D1-D5-P black hole}

In this section we describe the D1-D5-P BH and its thermodynamic
properties that are used in sections
\ref{sec:QSRflat}-\ref{sec:Extreme}. The most general solution with
arbitrary charges was originally constructed in \cite{Cvetic:1996xz}
(see also \cite{Giusto:2004id}). This solution generalizes the case
with equal D1 and D5 charges found previously in
\cite{Breckenridge:1996sn} and whose BPS limit yields the BMPV BH
\cite{Breckenridge:1996is}. Here we follow the notation of
\cite{Giusto:2004id,Jejjala:2005yu}.

This  three-charged BH is a solution of type II$B$ supergravity. The
only II$B$ fields that are turned on are the graviton $g_{\mu\nu}$,
the dilaton $\Psi$, and the RR 2-form $C\equiv C_{(2)}$. For the
field strength one has simply $F_{(3)}=dC_{(2)}$ since the RR field
$C_{(0)}$ and the NSNS field $H_{(3)}$ are absent. The type II$B$
action, in the Einstein frame, reduces in these conditions to
\bea
 I=\frac{1}{16\pi G_{10}}\int d^{10}x \sqrt{-g} \left[
  R-\frac{1}{2}\partial_{\mu}\Psi \partial^{\mu}\Psi- \frac{1}{12}
e^{\Psi}F_{(3)}^2 \right] \,,
 \label{S2bEinst}
  \eea
where $g$ is the determinant of the Einstein metric. The field
equations that follow from variation of action (\ref{S2bEinst}) are
\bea & & R_{\mu\nu}-\frac{1}{2}g_{\mu\nu}R -\frac{1}{2}
 \left( \partial_{\mu}\Psi \partial_{\nu}\Psi-
 \frac{1}{2}g_{\mu\nu}\partial_{\sigma}\Psi \partial^{\sigma}\Psi
 \right)
-\frac{1}{12} e^{\Psi}\left( 3 F_{\mu \alpha\beta}
F_{\nu}^{\:\:\alpha\beta}
- \frac{1}{2}g_{\mu\nu} F_3^2 \right)=0 \,, \nonumber \\
& & \frac{1}{\sqrt{-g}}\, \partial_{\mu}\left( \sqrt{-g} \,
g^{\mu\nu}
\partial_{\nu}\Psi \right)- \frac{1}{12}
e^{\Psi}F_3^2 =0 \,, \nonumber \\
& & \frac{1}{\sqrt{-g}}\,
\partial_{\mu}\left( \sqrt{-g} \, e^{\Psi} F^{\mu \alpha \beta} \right)=0 \,.
 \label{FieldEqs}
  \eea
Contraction of the graviton field equation yields for the Ricci
scalar,
\bea R=\frac{1}{2}\partial_{\mu}\Psi \partial^{\mu}\Psi+
\frac{1}{24} e^{\Psi}F_{3}^2 \,.
 \label{ricci}
 \eea
The graviton in the Einstein frame is (the relation between the
parameters describing the solution and the conserved charges is
displayed in (\ref{ADMcharges}))
\bea \label{3charge} ds^2&=&-\frac{f}{
\tilde{H}_{1}^{3/4}\tilde{H}_{5}^{1/4} }
 (dt^2 - dy^2)
 +\frac{M}{ \tilde{H}_{1}^{3/4}\tilde{H}_{5}^{1/4} } (s_p dy - c_p
dt)^2  \\
 & & + \tilde{H}_{1}^{1/4}\tilde{H}_{5}^{3/4}
\left(\frac{ r^2 dr^2}{ (r^2+a_{1}^2)(r^2+a_2^2) - Mr^2}
+d\theta^2 \right)\nonumber \\
 & & +  \frac{\tilde{H}_{1}
\tilde{H}_{5} - (a_2^2-a_1^2) ( \tilde{H}_{1} + \tilde{H}_{5} -f)
\cos^2\theta}{ \tilde{H}_{1}^{3/4}\tilde{H}_{5}^{1/4} } \, \cos^2
\theta \,d \psi^2 \nonumber \\
 & &  +  \frac{\tilde{H}_{1} \tilde{H}_{5}
    + (a_2^2-a_1^2) ( \tilde{H}_{1} + \tilde{H}_{5} -f) \sin^2\theta}
    { \tilde{H}_{1}^{3/4}\tilde{H}_{5}^{1/4} }
    \, \sin^2\theta \,d \phi^2
 + \frac{M}{ \tilde{H}_{1}^{3/4}\tilde{H}_{5}^{1/4} }\left(a_1 \cos^2 \theta
d \psi + a_2 \sin^2 \theta d \phi \right)^2 \nonumber \\
 & &
 + \frac{2M \cos^2 \theta}{ \tilde{H}_{1}^{3/4}\tilde{H}_{5}^{1/4} }
 {\biggl [}(a_1 c_1 c_5 c_p -a_2 s_1 s_5 s_p) dt + (a_2 s_1
s_5 c_p - a_1 c_1 c_5 s_p) dy {\biggr ]} d\psi \nonumber \\
 & &
 +\frac{2M \sin^2 \theta}{ \tilde{H}_{1}^{3/4}\tilde{H}_{5}^{1/4} }
 {\biggl [}(a_2 c_1 c_5 c_p - a_1 s_1 s_5 s_p) dt + (a_1 s_1 s_5 c_p - a_2 c_1 c_5
s_p ) dy {\biggr ]} d\phi
 + \tilde{H}_{1}^{1/4}\tilde{H}_{5}^{-1/4} \sum_{j=1}^4
dz_j^2 \,, \nonumber \eea
where $y$ is the coordinate on $S^1$, and $z_j$'s ($j=1,\cdots,4$)
are the coordinates on the torus $T^4$. We use the notation $c_i
\equiv \cosh \delta_i$, $s_i \equiv \sinh \delta_i$, and
\bea f(r)&=&r^2+a_1^2\sin^2\theta+a_2^2\cos^2\theta \,, \qquad
\tilde{H}_i(r)=f(r)+M s_i^2\,, \:\,{\rm with}\:\, i=1,5 \,,\nonumber \\
g(r)&=&(r^2+a_1^2)(r^2+a_2^2)-M r^2\,.
 \label{def f H}
 \eea
The roots of $g(r)$, $r_+$ and $r_-$, are given by \bea
 r_{\pm}^2= \frac{1}{2}\,(M-a_1^2-a_2^2) \pm \frac{1}{2}
 \sqrt{(M-a_1^2-a_2^2)^2-4a_1^2 a_2^2}\,, \label{roots}
\eea The system describes a regular BH\footnote{For $r_+^2<0$, {\it
i.e.}, $M \leq(a_1-a_2)^2$ the system can describe a smooth soliton
without horizon \cite{Jejjala:2005yu,Cardoso:2005gj}. We will not
discuss this solution.} when $r_+^2>0$, {\it i.e.}, for $M\geq
(a_1+a_2)^2$. The ten-dimensional determinant in the Einstein frame
is $\sqrt{-g}=r\sin\theta\cos\theta {\tilde H_1}^{1/4}{\tilde
H_5}^{3/4}$.
%
%
The dilaton  $\Psi$ and 2-form RR gauge potential $C$ which support
the D1-D5-P configuration are
\bea e^{2\Psi} = \frac{\tilde H_1}{\tilde H_5}\,, \label{dilaton}
\eea
\bea C_{(2)} &=& \frac{M }{\tilde H_1} {\biggl [} \cos^2 \theta
\left( a_{t\psi} dt + a_{y\psi} dy \right) \wedge d\psi
   + \sin^2 \theta
   \left( a_{t\phi} dt  + a_{y\phi} dy \right) \wedge d \phi
  \nonumber \\
& &\hspace{1cm} - s_1 c_1 dt \wedge dy -
  s_5 c_5 (r^2 + a_2^2 + M
  s_1^2) \cos^2 \theta \,d\psi \wedge d\phi{\biggr ]} \,, \label{rr}
\eea
where we defined
\bea a_{t\phi} &=& a_1 c_1 s_5 c_p - a_2 s_1 c_5 s_p \,, \qquad
 a_{t\psi}= a_2 c_1 s_5 c_p - a_1 s_1 c_5 s_p \,, \nonumber \\
a_{y\phi} &=& a_2 s_1 c_5 c_p - a_1 c_1 s_5 s_p \,, \qquad
 a_{y\psi}= a_1 s_1 c_5 c_p - a_2 c_1 s_5 s_p \,.
\label{auxa} \eea
By electric-magnetic duality\footnote{We use the convention
$\epsilon^{tr\theta\phi\psi y z_1 z_2z_3z_4}=1$, and the relation
(\ref{emduality}) is valid in the Einstein frame.},
\bea e^{\Psi}\sqrt{-g} F_{(3)}^{\mu_1\mu_2\mu_3}=\frac{1}{7!}
\epsilon^{\mu_1\mu_2\mu_3\nu_1 \cdots\nu_7}F^{(7)}_{\nu_1
\cdots\nu_7} \,, \label{emduality} \eea
our configuration can be equivalently  described either by the
2-form $C_{(2)}$ in (\ref{rr}) or by the 6-form $C_{(6)}$ that
follows from (\ref{emduality}). Using this equivalence, we rewrite
(\ref{rr}) as
\bea C_{(2)} &=& -\frac{M}{\tilde H_1}\left( s_1 c_1 dt + a_{y\phi}
\,\sin^2\theta d\phi +a_{y\psi} \,\cos^2\theta d\psi \right)
   \wedge dy \,,\nonumber \\
C_{(6)} &=& -\frac{M}{\tilde H_5}\left( s_5 c_5 dt + a_{t\psi}
\,\sin^2\theta d\phi
 +a_{t\phi} \,\cos^2\theta d\psi \right)
   \wedge dy\wedge dz_1\wedge dz_2\wedge dz_3\wedge dz_4  \,. \nonumber \\
& &  \label{rr2} \eea
The advantage of (\ref{rr2}) is that we clearly identify  the
$C_{(2)}$ gauge potential sourced by the D1-brane charges and the
$C_{(6)}$ field sourced by the D5-brane charges. Thus, this
expression will be appropriate to find the electric potentials
associated with the two type of D-branes. Note that all the
$C_{\mu\nu}^{(2)}$ components contain the $y$-coordinate that
parametrizes the $S^1$, while all the
$C_{\mu\nu\alpha\beta\gamma\sigma}^{(6)}$ components contain the
$y$-coordinate
 and the $z_j$'s  coordinates that parametrize the torus $T^4$.
This reflects the fact that D1-branes wrap $S^1$ and the D5-branes
wrap the full internal space $T^4\times S^1$.

The general procedure to compute angular velocities when the
geometry has several momenta can be found in \cite{Gibbons:2004uw}.
Applied to our case, the angular velocities at the horizon along the
$\phi$-plane, $\Omega^{\phi}$, the $\psi$-plane, $\Omega^{\psi}$,
and the velocity along $y$, $\Phi^{(p)}$ are\footnote{We identify
$\Omega^{y}\equiv \Phi^{(p)}$ because the KK momentum plays
effectively the role of a gauge charge with associated electric
potential.}
\bea \Omega^{\phi} &=& \frac{ g_{ty} \left(
g_{y\phi}g_{\psi\psi}-g_{y\psi}g_{\phi\psi} \right)
                      +g_{t\phi} \left( g_{y\psi}^2- g_{yy}g_{\psi\psi} \right)
                      +g_{t\psi} \left( g_{yy}g_{\phi\psi}-g_{y\phi}g_{y\psi} \right)  }
      { g_{yy}g_{\phi\phi}g_{\psi\psi}+2g_{y\phi}g_{y\psi}g_{\phi\psi}
       -g_{y\psi}^2g_{\phi\phi}-g_{y\phi}^2g_{\psi\psi}-g_{\phi\psi}^2g_{yy} }{\biggl |}_{r=r_+} \,, \nonumber \\
\Omega^{\psi} &=& \frac{ g_{ty} \left(
g_{y\psi}g_{\phi\phi}-g_{y\phi}g_{\phi\psi} \right)
                      +g_{t\phi} \left( g_{yy}g_{\phi\psi}-g_{y\phi}g_{y\psi} \right)
                      +g_{t\psi} \left( g_{y\phi}^2- g_{yy}g_{\phi\phi}  \right)  }
   { g_{yy}g_{\phi\phi}g_{\psi\psi}+2g_{y\phi}g_{y\psi}g_{\phi\psi}
       -g_{y\psi}^2g_{\phi\phi}-g_{y\phi}^2g_{\psi\psi}-g_{\phi\psi}^2g_{yy} }{\biggl |}_{r=r_+} \,,\nonumber \\
\Phi^{(p)} &=&  \frac{ g_{ty} \left( g_{\phi\psi}^2-
g_{\phi\phi}g_{\psi\psi} \right) +g_{t\phi}\left(
g_{y\phi}g_{\psi\psi}-g_{y\psi}g_{\phi\psi} \right)
        +g_{t\psi}\left( g_{y\psi}g_{\phi\phi}-g_{y\phi}g_{\phi\psi} \right)  }
        { g_{yy}g_{\phi\phi}g_{\psi\psi}+2g_{y\phi}g_{y\psi}g_{\phi\psi}
       -g_{y\psi}^2g_{\phi\phi}-g_{y\phi}^2g_{\psi\psi}-g_{\phi\psi}^2g_{yy} }{\biggl |}_{r=r_+}
       \,,\nonumber \\
    & &   \label{defVelocity}
\eea
%
%
%
%
which yields
\bea \Omega^{\phi} &=& -\frac{ a_2 r_+^2}
 {\left( r_+^2+a_2^2 \right) \left( r_+^2c_1c_5c_p+a_1a_2s_1s_5s_p \right)}
 \,, \nonumber \\
\Omega^{\psi} &=& -\frac{ a_1 r_+^2 }
 {\left( r_+^2+a_1^2 \right) \left( r_+^2c_1c_5c_p+a_1a_2s_1s_5s_p \right) }  \,,\nonumber \\
\Phi^{(p)} &=&  \frac{ r_+^2c_1c_5s_p+a_1a_2s_1s_5c_p }
                 { r_+^2c_1c_5c_p+a_1a_2s_1s_5s_p }  \,. \label{Velocity}
\eea
The horizon angular velocities are constant and, in particular have
no angular dependence, as required by Carter's rigidity property of
Killing horizons\footnote{Note that the angular velocities can be
more easily computed using the standard formulas valid for solutions
rotating along a single axis, as long as we evaluate them at a
specific $\theta$ coordinate. More concretely, an inspection of
(\ref{defVelocity}) concludes that the following relations are valid
and provide the quickest computation of the corresponding
quantities: \bea \Omega^{\phi}\equiv
\frac{g_{t\phi}}{g_{\phi\phi}}{\biggl |}_{r=r_+,\,\theta=0}\,,\qquad
\Omega^{\psi}\equiv \frac{g_{t\psi}}{g_{\psi\psi}}{\biggl
|}_{r=r_+,\,\theta=\pi/2}\,,\qquad \Phi^{(p)}\equiv
\frac{g_{ty}}{g_{yy}}{\biggl |}_{r=r_+,\,\theta=0 }\,. \nn \eea
 }.
%
%
%
%
The electric potentials at the horizon associated with the $Q_1$ and
$Q_5$ gauge charges are computed using
\bea
\Phi^{(i)} &=& -C_{\mu\{x^{(i)}\} } \xi^{\mu}{\biggl |}_{r=r_+} \nonumber \\
 &=&  -\left[ C_{t\{x^{(i)}\} }+C_{\phi\{x^{(i)}\} } \Omega^{\phi}
  +C_{\psi\{x^{(i)}\} } \Omega^{\psi}\right]_{r=r_+}\,, \qquad i=1,5 \,,
  \label{def:electricpotential}
\eea
%
%
%
where,
\bea \xi=\partial_t+\Omega^{\phi}\partial_{\phi}
+\Omega^{\psi}\partial_{\psi} \,, \label{killing} \eea
is the null Killing vector generator of the horizon
($\Omega^{\phi,\psi}$ are the horizon angular velocities). We use
the notation $\{x^{(1)}\}\equiv y$, the coordinate of $S^1$ wrapped
by D1-branes, and $\{x^{(5)}\}\equiv yz_1z_2z_3z_4$, the coordinates
of $S^1\times T^4$  wrapped by D5-branes. Using the $C$ gauge
potential written in (\ref{rr2}) yields
\bea \Phi^{(i)} = \frac{ r_+^2(\tanh\delta_i)c_1c_5c_p +
a_1a_2(\coth\delta_i)s_1s_5s_p }
                 {r_+^2c_1c_5c_p+a_1a_2s_1s_5s_p} \,, \qquad i=1,5  \,.
         \label{electricpotential}
\eea
The temperature of the BH is $T=\kappa_{\rm h}/(2\pi)$ where the
surface gravity of the horizon is $\kappa_{\rm
h}^2=-\frac{1}{2}(\nabla^{\mu}\xi^{\nu})(\nabla_{\mu}\xi_{\nu})
 {\bigl |}_{r=r_+}$, and $\xi^{\mu}$ is the Killing vector
horizon generator defined in (\ref{killing}). The inverse
temperature $\beta=1/T$ is then
\bea \beta
=\frac{2\pi\left(r_+^2+a_1^2\right)\left(r_+^2+a_2^2\right)}{r_+\left(r_+^4-a_1^2a_2^2\right)}
 \left(r_+^2c_1c_5c_p+a_1a_2s_1s_5s_p\right)  \,. \label{beta}
\eea
The entropy $S$ is just horizon area (in the Einstein frame) divided
by $4G_{10}$,
\bea
S=\frac{2\pi\left(r_+^2+a_1^2\right)\left(r_+^2+a_2^2\right)}{r_+^3}
 \left(r_+^2c_1c_5c_p+a_1a_2s_1s_5s_p\right)  \,. \label{entropy}
\eea

To conclude this section, note that action (\ref{S2bEinst}) can be
written in the string frame through the Weyl rescaling of the
metric, $\tilde{g}_{AB}=e^{\Psi/2} g_{AB}$, yielding
\bea
 I=\frac{1}{16\pi G_{10}}\int d^{10}x \sqrt{-\tilde{g}} \left[e^{-2\Psi}\left(
  \tilde{R}-4\partial_{\mu}\Psi \partial^{\mu}\Psi\right)- \frac{1}{2\cdot 3!}
F_{(3)}^2 \right] \,.
 \label{SIIbstring}
  \eea
%

\subsection{\label{sec:nearBPS} The near-BPS limit of the D1-D5-P black hole}

 In this appendix we present the
detailed computation of the near-BPS limit of the D1-D5-P BH, and of
the off-BPS construction that takes
(\ref{betaLR})-(\ref{electricpotent(M)}) into
(\ref{expand:betaLR})-(\ref{SQSR:d1d5p}).

Using the trignometric properties
\bea c_i = \frac{e^{\delta_i}+e^{-\delta_i}}{2}\,, \qquad s_i =
\frac{e^{\delta_i}-e^{-\delta_i}}{2}\,, \label{trig} \eea
the gauge charges and ADM mass (\ref{ADMcharges}) are, in the
near-BPS regime (\ref{nearBPSlim}),
\bea Q_p &=& \frac{M e^{2\delta_p}}{4}-\frac{M e^{-2\delta_p}}{4}
\equiv Q_p^{bps}-\varepsilon \nonumber \\
Q_1 &\simeq & \frac{M e^{2\delta_1}}{4}\equiv Q_1^{bps}\,,
 \qquad Q_5 \simeq  \frac{M e^{2\delta_5}}{4}\equiv Q_5^{bps}\,, \nonumber \\
E &\simeq & \left(Q^{bps}_1 +Q^{bps}_5+ Q^{bps}_p
\right)+\varepsilon
     =E^{bps}+\varepsilon \,, \label{ADMmovers}
\eea
where the BPS constraint (\ref{ADMchargesBPS}) was used. We can
interpret the quantity $\frac{M e^{2\delta_p}}{4}$ as the number of
left-movers, and $\varepsilon=\frac{M e^{-2\delta_p}}{4}$ as the
number of right-movers (in the KK momentum sector). The BPS
configuration, $\varepsilon=0$, is the one with no right-movers. In
the D1 and D5 sectors there are only left-movers since
$\delta_{1,5}\rightarrow\infty$. From the last relation in
(\ref{ADMmovers}), we conclude that $\varepsilon$ is also an off-BPS
parameter that measures energy above extremality. We can also
rewrite $\varepsilon=\frac{M e^{-2\delta_p}}{4}$ as
$M=4\sqrt{Q_p^{bps}}\sqrt{\epsilon}$, an expression that will be
useful below.

The near-BPS limit (\ref{nearBPSlim}) is completed with the angular
momenta condition. It can be understood as follows. Inversion of
(\ref{ADMcharges}) yields
\bea a_1 =  -\frac{1}{M}\frac{J_{\psi}c_1c_5c_p +
J_{\phi}s_1s_5s_p}{c_1^2c_5^2c_p^2 - s_1^2 s_5^2 s_p^2}   \,,\qquad
a_2 =  -\frac{1}{M}\frac{J_{\phi}c_1c_5c_p + J_{\psi}s_1s_5s_p }
       {c_1^2c_5^2c_p^2 - s_1^2 s_5^2 s_p^2}    \,.
         \label{a1a2}
\eea
In the near-BPS limit (\ref{nearBPSlim}) one has $c_{1,5}^2 \simeq
\frac{Q_{1,5}^{\rm bps}}{M}+\frac{1}{2}$, $s_{1,5}^2 \simeq
\frac{Q_{1,5}^{\rm bps}}{M}-\frac{1}{2}$, $c_{p}^2 = \frac{Q_p^{\rm
bps}}{M}+\frac{1}{2}+\frac{\varepsilon}{M}$, and $s_{p}^2 =
\frac{Q_p^{\rm bps}}{M}-\frac{1}{2}+\frac{\varepsilon}{M}$. The
expansion of $a_{1,2}$ in the small $M$ regime then gives
\bea a_{1,2} =  -\left( J_{\phi}+J_{\psi} \right)
\frac{\sqrt{\gamma}}{\eta}\frac{1}{\sqrt{M}} \mp \frac{1}{4} \left(
J_{\phi}-J_{\psi} \right) \frac{ \sqrt{M} }{\sqrt{\gamma}}
+\mathcal{O}\left( M^{3/2} \right) \,,
         \label{expand:a1a2}
\eea
where we have defined $\eta\equiv Q^{bps}_1 Q^{bps}_5+Q^{bps}_1
Q^{bps}_p+Q^{bps}_5 Q^{bps}_p + \left(Q^{bps}_1 +Q^{bps}_5\right)
\varepsilon$, and $\gamma\equiv Q^{bps}_1 Q^{bps}_5 Q^{bps}_p
+Q^{bps}_1 Q^{bps}_5 \varepsilon$. Now, one must take appropriate
limits of $a_1$ and $a_2$ such that they keep finite and the angular
momenta is kept fixed. But in (\ref{expand:a1a2}) one sees that, for
non-vanishing charges $Q^{bps}_i\neq 0$ ($i=1,5,p$), $a_{1,2}$
diverge as $1/\sqrt{M}$ when we take $M\rightarrow 0$. We can avoid
this divergence by imposing that $J_{\phi}+J_{\psi}\rightarrow 0$ in
the near-BPS limit. Note that as a consequence, in the limit
$\varepsilon \rightarrow 0$, the BPS solution must have angular
momenta satisfying the relation
(\ref{ADMchargesBPS})\footnote{Alternatively, note that we could
relax this condition in the off-BPS regime. That is we could instead
fix $J_{\phi}$ and let $J_{\psi}$ arbitrary ``during'' the near-BPS
approach, as long as in the BPS limit one ended with
$J_{\phi}+J_{\psi}=0$. Our final result is independent of the
particular off-BPS path choosen.}. Under this condition, we can now
take a small $\epsilon$ expansion in (\ref{expand:a1a2}) and get
\bea a_{1,2} =  \pm \sqrt{M} \frac{J_{\phi}^{\rm
bps}}{2\sqrt{Q^{bps}_1
Q^{bps}_5 Q^{bps}_p}} \left[ 1 
+\mathcal{O}\left( \varepsilon^{} \right) \right] \,.
\label{expand:a1a2+} \eea

Use of (\ref{trig}) and (\ref{expand:a1a2+}) in (\ref{betaLR}),
(\ref{entropy(M)}) and (\ref{velocity(M)}) yields straightforwardly
the near-BPS expansions for the temperature, (\ref{expand:betaLR}),
for the entropy, (\ref{Sbps}), and for the angular velocities,
(\ref{expand:potent}), respectively.

The off-BPS expansion of the electric potential $\Phi^{(p)}$ leading
to (\ref{expand:potent}) is straightforward. However, the expansion
of the D1 and D5 electric potentials is more subtle. Indeed, if in
(\ref{electricpotent(M)}) we do the most natural step,
$(\tanh\delta_1)c_1c_5c_p - (\coth\delta_1)s_1s_5s_p = s_1c_5c_p -
c_1s_5s_p$ we just catch the BPS value but not the next order term
of the expansion. To capture the next order off-BPS contribution one
has to introduce the parameter $M$ that measures the energy above
extremality. This is consistently done with the following step:
$(\tanh\delta_1)c_1c_5c_p=c_1c_5c_p\frac{Ms_1c_1}{M(1+s_1^2)}$ (and
similarly for the term proportional to $\coth\delta_1$). Then, use
of $Ms_1^2\simeq Q_1^{\rm bps}-M/2$ and
$M=4\sqrt{Q_p^{bps}}\sqrt{\epsilon}$ allows  to finally write
$(\tanh\delta_1)c_1c_5c_p \simeq c_1c_5c_p(1-q \sqrt{\varepsilon})$,
where $q$ is a ratio of BPS charges. The expansion
(\ref{expand:potent}) for $\Phi^{(1)}$, $\Phi^{(5)}$ now follows
naturally.

\setcounter{equation}{0}\section{Explicit agreement for other black
hole systems} \label{sec:several}

In this Appendix we will perform the Euclidean zero temperature
limit and study the statistical mechanics of some BHs that have not
been considered in the main body of the text. The main motivation to
do this is two-folded. First, we explicitly verify that the relation
between the Euclidean zero temperature and Sen's entropy formalisms
is indeed general and not restricted to the three-charged BH studied
in the main body of the text.  Second, we get a list of conjugated
chemical potentials for several BH systems. With these at hand we
can also study the thermodynamics of the dual CFT. We consider some
relevant asymptotically flat systems that have been discussed within
Sen's formalism context in \cite{Astefanesei:2006dd}, namely: the
four-charged BH (subsection \ref{sec:4Q}), and the Kerr-Newman BH
(subsection \ref{sec:KN}). The agreement between the two formalisms
is also confirmed for black holes of gauged supergravity elsewhere
\cite{SQSR1,Silva:2007tw}.

\subsection{Four-charged black
holes} \label{sec:4Q}

We study the statistical properties at zero temperature of the
asymptotically flat four-charged BH in four dimensions ($4D$). This
system has three distinct extreme cases: the BPS BH (studied in
subsection \ref{sec:4Q:bps}), the ergo-free branch family of BHs
(subsection \ref{sec:4Q:ergofree}), and the ergo-branch family
(subsection \ref{sec:4Q:ergobranch}). These last two are extreme but
not BPS BHs and we are following the nomenclature of
\cite{Astefanesei:2006dd}.

The most general non-extremal rotating four-charged BH was first
found in \cite{Cvetic:1996kv} as a solution of heterotic string
theory compactified on a six-torus. The four gauge fields of the
solution were however not explicitly given. This BH is also a
solution of $N=2$ supergravity coupled to three vector multiplets,
which in turn can be consistently embedded in $N=8$ maximal
supergravity \cite{Cvetic:1996kv,Horowitz:1996ac,Chong:2004na}. As
first observed for the static non-extreme case
\cite{Horowitz:1996ac}, these theories can also be obtained from
compactification of type II supergravity on $T^4\times S^1 \times
\tilde{S}^1$. Therefore, from the $10D$ viewpoint these BHs have a
D-brane interpretation, {\it e.g.}, they describe the D2-D6-NS5-P
solution of type II$A$ supergravity or the D1-D5-KK-P solution of
type II$B$ supergravity (or any dual system to these obtained by
$U$-dualities).

Take $N=2$ supergravity coupled to three vector multiplets. The
field content of the theory is: the graviton $g_{\mu\nu}$, four
gauge fields $A_{\1 1,2}\,,\hat\cA_\1^{1,2}$, three dilatons
$\varphi_i$ and three axions $\chi_i$ (with $1\le i \le 3$). The
full solution can be explicitly found in \cite{Chong:2004na}.
Compared with \cite{Chong:2004na}, we use the parameters $\mu \equiv
4m$ and $l\equiv 4a$ that avoid nasty factors of $4$ in the
thermodynamic quantities. The horizons of the solution are at
\bea r_{\pm}= \frac{1}{4}\left( \mu \pm \sqrt{\mu^2-l^2} \right) \,,
\label{4Q:horizons} \eea
and thus the system has regular horizons when $ \mu \geq |l|$. When
$l=0$ we recover the static solutions found in
\cite{Horowitz:1996ac}.

The conserved mass $E$, angular momentum $J$, and gauge charges
$Q_i$'s of the BH are (we use  $G_4\equiv 1/8$ for this system)
\bea E &=& \frac{\mu}{2} \sum_{i=1}^{4} \cosh (2\delta_i)\,, \qquad
J_{\phi} = \frac{1}{2} \mu l \left(c_1 c_2 c_3 c_4
- s_1 s_2 s_3 s_4\right)\,,\nonumber \\
 Q_i &=& \mu s_i c_i\,, \qquad i=1,2,3,4 \,, \label{4Q:ADMcharges}
\eea
which are invariant under interchange of the $\delta_i$'s, as
expected from the $U$-duality relations.

 The left and right movers inverse
temperatures, the entropy, electric potentials and angular velocity
are \cite{Cvetic:1997xv},
\bea \beta_L &=&  2\pi \mu \left(c_1 c_2 c_3 c_4 - s_1 s_2 s_3
s_4\right) \,, \qquad \beta_R = \frac{ 2\pi \mu^2}
{\sqrt{\mu^2-l^2}}
                \left(c_1 c_2 c_3 c_4 + s_1 s_2 s_3 s_4\right)\,, \nonumber \\
S &=&   \pi \mu^2 \left(c_1 c_2 c_3 c_4 + s_1 s_2 s_3 s_4\right)
       + \pi \mu\sqrt{\mu^2-l^2} \left(c_1 c_2 c_3 c_4 - s_1 s_2 s_3 s_4\right) \,,
                                         \nonumber \\
\Phi^{(i)} &=&  \frac{\pi \mu} {\beta} \left[ (\tanh\delta_i)c_1
c_2 c_3 c_4 - (\coth\delta_i)s_1 s_2 s_3 s_4 \right] \nonumber \\
 & &+\frac{\pi \mu^2}
{\beta\sqrt{ \mu^2-l^2}} \left[ (\tanh\delta_i)c_1 c_2 c_3 c_4 +
(\coth\delta_i)s_1 s_2 s_3 s_4
\right], \quad i=1,2,3,4  \,, \nonumber \\
\Omega &=& \frac{1}{\beta}
              \frac{2\pi l}{\sqrt{ \mu^2-l^2}} \,.  \label{4Q:thermo}
\eea
%

\subsubsection{BPS black hole}
 \label{sec:4Q:bps}

The BPS limit of the four charged BH is obtained by taking $\mu
\rightarrow 0$, $\delta_i\rightarrow \infty$, while keeping $Q_i$
fixed ($i=1,2,3,4$), and $l\rightarrow 0$ at the same rate as $\mu$,
{\it i.e.}, $l/\mu \rightarrow 1$. As a consequence $J\rightarrow 0$
and the BPS four-charged BH is non-rotating\footnote{The reason
being that the roots that define the horizon are
(\ref{4Q:horizons}), and thus $\mu \geq |l|$ must hold to have a
regular solution.}. Therefore, the BPS charges satisfy the BPS
constraints,
\bea E^{bps}= Q_1^{bps}+Q_2^{bps}+Q_3^{bps}+Q_4^{bps} \,,  \qquad
J^{bps}=0  \,, \label{4Q:ADMchargesBPS} \eea
where $Q_i^{bps}=\mu e^{2\delta_i}/4$. To study the thermodynamics
near the $T=0$ BPS solution we work in the {\it near}-BPS limit. We
take
\bea \mu \rightarrow 0 \,, \quad \delta_{1,2,3} \rightarrow \infty
\,, \quad Q_{1,2,3} \:\:{\rm fixed}\,;  \quad
 \delta_4 \:\:{\rm finite}; \quad l\rightarrow 0 \:\:\:(l/\mu \rightarrow 1)\,. \label{4Q:nearBPSlim}
\eea
Note that we take the four boosts to be positive and we choose to
keep $\delta_4$ finite, without any loss of generality (due to
$U$-dualities).

Define the off-BPS parameter above extremality $\varepsilon$, to be
$\varepsilon= \mu  e^{-2\delta_4}/4$ so that $E\equiv E^{\rm
bps}+\varepsilon$. The procedure yielding the off-BPS expansion of
the several thermodynamic quantities is quite similar to the one
done in the three-charged BH (see Appendix \ref{sec:nearBPS}). So we
just quote the relevant results.

 Expanding the left and right temperatures in terms $\varepsilon$
 yields,
\bea \beta_L =  \pi \sqrt{\frac{Q^{bps}_1 Q^{bps}_2 Q^{bps}_3}
{Q^{bps}_4} }
 \,,\qquad
\beta_R = \pi \sqrt{Q^{bps}_1 Q^{bps}_2 Q^{bps}_3}\frac{1}{
\sqrt{\varepsilon} } \,. \label{4Q:expand:betaLR} \eea
The BPS limit corresponds to send $\beta_R \rightarrow \infty$, and
we now can use $\beta_R$ as the off-BPS parameter, instead of
$\varepsilon$.

The expansion in $\beta_R$ of the conserved charges is
\bea
& & E= E^{bps}
  +\mathcal{O}\left( \beta_R^{-2} \right) \,, \qquad
J = \frac{\pi Q^{bps}_1 Q^{bps}_2 Q^{bps}_3}{\beta_R}
  +\mathcal{O}\left( \beta_R^{-2} \right)
    \,,\nonumber \\
 & & Q_{1,2,3} \simeq Q_{1,2,3}^{bps}\,, \qquad
 Q_4 = Q_4^{bps} 
 +\mathcal{O}\left( \beta_R^{-2}\right)\,. \label{4Q:expand:ADMcharges}
\eea
The remaining thermodynamic quantities have the expansion,
\bea
S &=& S^{bps}
  +\mathcal{O}\left( \beta_R^{-1} \right) \,, \qquad
\Omega=
 \frac{4\pi }{\beta_R} +\mathcal{O}\left( \beta_R^{-2}\right) \,,
                                       \nonumber\\
\Phi^{(i)} &=& \Phi_{bps}^{(i)}-\frac{2 \phi_{i} }{\beta_R}
+\mathcal{O}\left( \beta_R^{-2}\right)\,, \qquad i=1,2,3,4 \,,
   \label{4Q:expand:potent}
\eea
where
\bea S^{bps}&=& 2\pi \left[ Q^{bps}_1 Q^{bps}_2 Q^{bps}_3 Q^{bps}_4
\right]^{1/2}
                  \,, \nonumber \\
 \Phi_{bps}^{(i)}&=& 1\,, \qquad
   \phi_{i} = \frac{\pi \left[ Q^{bps}_1 Q^{bps}_2 Q^{bps}_3 Q^{bps}_4 \right]^{1/2}}
                     { Q_i^{bps} } \,, \qquad
i=1,2,3,4 \,.
    \label{4Q:ConjPotential}
\eea
The last relation gives the key quantities, namely the conjugated
potentials $\phi_{i}$'s of the solution that have an important role
in the dual CFT. The expressions of the BPS entropy $S_{bps}$, and
conjugated potentials  $\phi_i$'s agree with the corresponding
quantities computed in \cite{Astefanesei:2006dd} using Sen's entropy
function formalism\footnote{\label{4Q:foot}Once we match the
notation $\omega\equiv 2\pi \alpha$ and $\phi_i\equiv 2\pi e_i$ and
we take into consideration that we use $G_4\equiv 1/8$, while
\cite{Astefanesei:2006dd} uses $G_4\equiv 1/(16\pi)$).}.

The SQSR for the four-charge BH is then
\bea I_{bps}=\phi_1\, Q_1^{bps}+\phi_2\, Q_2^{bps}+\phi_3\,
Q_3^{bps} +\phi_4\, Q_4^{bps} -S_{bps}\,.
 \label{4Q:SQSR}
\eea
%

\subsubsection{Extreme (non-BPS) black hole: ergo-free solution}
\label{sec:4Q:ergofree}

In the four-charged system we can take an extremal limit that yields
a rotating BH without ergosphere. For this reason, this BH was
dubbed ergo-free solution in \cite{Astefanesei:2006dd}.

This limit is similar to the BPS regime token in the previous
Appendix \ref{sec:4Q:bps} in which we send the boosts to infinity;
the difference being that we take an odd number (one, for
definiteness, but it could as well be three) of boosts to be
negative. As explained in a similar context in section
\ref{sec:Extreme}, this limit yields an extreme, but not BPS, BH.

Concretely, take the {\it near}-extremal limit ($\delta_{1,2,3}>0$;
$\delta_4<0$, $Q_4<0$):
\bea \mu \rightarrow 0 \,, \quad \delta_{1,2,3} \rightarrow \infty
\,, \quad Q_{1,2,3} \:\:{\rm fixed}\,;  \quad
 \delta_4<0 \:\:{\rm finite}\,;
 \quad \frac{l}{\mu}\rightarrow \frac{J}{\sqrt{-Q_1Q_2Q_3Q_4}}\,.
 \label{free:nearBPSlim}
\eea
The charges in the extreme solution satisfy the constraint
\bea E^{ext}= Q_1^{ext}+Q_2^{ext}+Q_3^{ext}-Q_4^{ext} \,,
 \label{free:ADMchargesBPS}
\eea
where $Q_{1,2,3}^{bps}=\mu e^{2\delta_{1,2,3}}/4$, $Q_4^{ext}=-\mu
e^{-2\delta_4}/4$, and $J^{ext}$ is arbitrary. Using the
off-extremality parameter, $\varepsilon=\mu e^{2\delta_4}/4=\pi^2
Q_1^{ext}Q_2^{ext}Q_3^{ext}\beta_L^{-2}$ (so the extremal limit is
obtained by sending $\beta_L \rightarrow \infty$), we get the
following expansion for the relevant thermodynamic quantities:
\bea
& & S = S_{ext} 
  +\mathcal{O}\left( \beta_L^{-1} \right) \,, \qquad  \Omega=\Omega_{ext}-
 \frac{2\omega}{\beta_L}
                +\mathcal{O}\left( \beta^{-2}\right) \,,
 \nonumber \\
& & \Phi^{(i)} = \Phi_{ext}^{(i)}-\frac{2 \phi_{i} }{\beta_L}
+\mathcal{O}\left( \beta_L^{-2}\right)\,, \qquad i=1,2,3,4 \,,
   \label{free:expand:potent}
\eea
where
\bea & & S_{ext}= 2\pi \left[ -Q^{ext}_1 Q^{ext}_2 Q^{ext}_3
Q^{ext}_4-(J^{ext})^2 \right]^{1/2}
                  \,, \nonumber \\
& &  \Omega_{ext}=0\,, \qquad \Phi_{ext}^{(1,2,3)}= 1\,, \qquad
\Phi_{ext}^{(4)}= -1\,.
    \label{free:ext}
\eea
The conjugated potentials are
\bea &&\omega = -\frac{2\pi J^{ext}}
{\left[ -Q^{ext}_1 Q^{ext}_2 Q^{ext}_3 Q^{ext}_4-(J^{ext})^2 \right]^{1/2}} \nonumber\\
&& \phi_{i} = -\frac{\pi  Q^{ext}_1 Q^{ext}_2 Q^{ext}_3 Q^{ext}_4 }
  { Q_i^{ext}\left[ -Q^{ext}_1 Q^{ext}_2 Q^{ext}_3 Q^{ext}_4-(J^{ext})^2 \right]^{1/2} } \,, \qquad
i=1,2,3,4 \,.
    \label{free:ConjPotential}
\eea
Again, these expressions for $S_{ext}$, $\omega$ and $\phi_i$'s
match the ones found in \cite{Astefanesei:2006dd} using Sen's
entropy function formalism  (see footnote \ref{4Q:foot} for
normalization conventions).

Although this is a non-BPS solution, it satisfies the extremal
constraint (\ref{free:ADMchargesBPS}) that is linear in the charges.
Using in addition (\ref{free:ext}), we find that (\ref{Ext:split2}),
applied to this system, vanishes and the QSR for this system
simplifies to
\bea I_{ext}=\sum_{i=1}^{4}\phi_i\, Q_i^{ext}+\omega J^{ext}
-S_{ext}\,.
 \label{free:QSR}
\eea
This is an example of a rotating extreme solution without
ergosphere. It has a finite on-shell action.

\subsubsection{Extreme (non-BPS) black hole: ergo-branch solution}
\label{sec:4Q:ergobranch}

This time we take the limit $\mu \rightarrow l$. This yields an
extreme BH with an ergosphere that was coined as ergo-branch
solution in \cite{Astefanesei:2006dd} (This is the four-charged
counterpart of the solution studied in Section
\ref{sec:Extreme:ergo3Q}).

We take the near-extreme limit
\bea \mu \rightarrow l+\varepsilon \,, \quad \varepsilon\ll 1 \,.
\label{ergo:nearEXTlim} \eea
When the off-extreme parameter $\varepsilon$ vanishes, the
temperature  indeed vanishes since $\beta_R \rightarrow \infty$ in
(\ref{4Q:thermo}). The off-extreme expansion of the conserved
charges (\ref{4Q:ADMcharges}) around the corresponding extreme
values (obtained by replacing $\mu$ by $l$ in (\ref{4Q:ADMcharges}))
is straightforward, and the expansion of the thermodynamic
quantities (\ref{4Q:thermo}) yields\footnote{We use the relation
$\sqrt{\mu^2-l^2}\simeq 2\pi \mu^2 (c_1 c_2 c_3 c_4 +s_1 s_2 s_3
s_4)/\beta_R$}
\bea & & \beta_L =  2\pi l \left(c_1 c_2 c_3 c_4 - s_1 s_2 s_3
s_4\right)
       +\mathcal{O}\left( \varepsilon \right)
 \,,\qquad
\beta_R = \sqrt{2} \pi l^{3/2} \left(c_1 c_2 c_3 c_4+ s_1 s_2 s_3
s_4\right) \frac{1}{ \sqrt{\varepsilon} } \,, \nonumber \\
& & S = S_{ext}
  +\mathcal{O}\left( \beta_R^{-1} \right) \,, \qquad
\Omega= \Omega_{ext}-\frac{2\omega}{\beta_R} +\mathcal{O}\left(
\beta_R^{-2}\right) \,,
                                       \nonumber\\
& & \Phi^{(i)} = \Phi_{ext}^{(i)}-\frac{2 \phi_{i} }{\beta_R}
+\mathcal{O}\left( \beta_R^{-2}\right)\,, \qquad i=1,2,3,4 \,,
   \label{Ergo:expand}
\eea
where the extreme values satisfy
\bea & & S_{ext}= 2\pi \left[ Q^{ext}_1 Q^{ext}_2 Q^{ext}_3
Q^{ext}_4 +(J^{ext})^2 \right]^{1/2}
                  \,, \qquad
  \Omega_{ext}= 2 l^{-1} \left( c_1 c_2 c_3 c_4 +s_1 s_2 s_3 s_4 \right)^{-1}
                                   \,, \nonumber \\
 & & \Phi_{ext}^{(i)}= \frac{
       (\tanh\delta_i)c_1 c_2 c_3 c_4 + (\coth\delta_i)s_1 s_2 s_3 s_4}
     {c_1 c_2 c_3 c_4 +s_1 s_2 s_3 s_4} \,,  \quad i=1,2,3,4       \,,
    \label{Ergo:extS}
\eea
and the conjugated potentials are
\bea
 & & \omega= \frac{2\pi J^{ext}}
 { \left[ Q^{ext}_1 Q^{ext}_2 Q^{ext}_3 Q^{ext}_4 +(J^{ext})^2\right]^{1/2} }
                                   \,, \nonumber \\
 & &  \phi_{i} = \frac{2 \pi l^2}{Q^{ext}_i} \frac{s_1c_1 s_2c_2 s_3c_3 s_4c_4}
 {c_1 c_2 c_3 c_4 +s_1 s_2 s_3 s_4}  \,, \qquad
i=1,2,3,4 \,.
    \label{Ergo:conjPot}
\eea
Note that in the last expression could be rewritten only in terms of
the conserved charges as expected by the attractor mechanism. We do
not do it here because the expression is too long. The expressions
of the extremal entropy $S_{ext}$, and conjugated potentials
$\omega$ and $\phi_i$'s agree with the corresponding quantities
computed in \cite{Astefanesei:2006dd} using Sen's entropy function
formalism (see footnote \ref{4Q:foot} for normalization
conventions).

The QSR for this system is
\bea  I= \beta\left( E^{ext}-\sum_{i=1}^{4} \Phi_{ext}Q_i^{ext}
  -\Omega_{ext}J^{ext}\right)+
\sum_{i=1}^{4}\phi_i\, Q_i^{ext}+\omega\, J^{ext} -S_{ext}
+\mathcal{O}\left( \beta_R^{-1}\right)\,
 \label{Ergo:QSR}
\eea
In the supersymmetric system the analogue of the first term vanishes
due to the BPS constraint on the conserved charges. But, in general,
for non-BPS extreme BHs it does not vanish (see also discussion
associated with (\ref{Ext:split2})). In the present case the factor
in between brackets is $-\frac{l}{2}\frac{c_1 c_2 c_3 c_4 -s_1 s_2
s_3 s_4}{c_1 c_2 c_3 c_4 +s_1 s_2 s_3 s_4}$. Note that this quantity
vanishes when rotation is absent. When it is present, the solution
has an ergosphere and the non-vanishing contribution seems to be
associated with its existence, as discussed in section
\ref{sec:Extreme}.

\subsection{Extreme Kerr-Newman black hole}
 \label{sec:KN}

In this section we take the near-extreme limit of the Kerr-Newman BH
with ADM mass $M$, ADM charge $Q$ and ADM angular momentum $J=aM$
that is a solution of the Einstein-Maxwell action
$I=\frac{1}{16\pi}\int d^4x\sqrt{-g}\left(R-F^2\right)$ (so, we set
$G_4\equiv 1$). In the extreme state the charges satisfy the
constraint $M^2=a^2+Q^2$, the horizons coincide, $r_{\pm}=M$, and
one also has the useful relation $M^2+a^2=2\sqrt{J^2+Q^4/4}$. Define
the off-extremality parameter $\varepsilon$ such that
$M=M_{e}+\varepsilon$ which implies that $r_+\sim
M_{e}+\sqrt{2M_{e}}\sqrt{\varepsilon}$ (the subscript $e$ stands for
the on-shell extreme solution). In terms of the inverse temperature
$\beta=\frac{2\pi (r_+^2+a^2)}{r_+-M}$ it is given by
$\sqrt{\varepsilon}=\frac{2\pi(M_{e}^2+a_e^2)}{\sqrt{2M_e}\,\beta}$.
Using the expressions $S=\pi(r_+^2+a^2)$, $\Omega=a/(r_+^2+a^2)$ and
$\Phi=Qr_+/(r_+^2+a^2)$ one gets the expansion:
\bea & & S = S_{e}+\mathcal{O}\left( \beta^{-1} \right),
         \qquad S_e=2\pi \sqrt{J_e^2+Q_e^4/4} \,; \nonumber \\
& & \Omega=\Omega_{e}-
 \frac{\omega}{\beta}+\mathcal{O}\left( \beta^{-2}\right)\,,
 \qquad \Omega_e=\frac{J_e}{2M_e\sqrt{J_e^2+Q_e^4/4}} \,,
 \qquad \omega=\frac{2\pi J_e}{\sqrt{J_e^2+Q_e^4/4}}\,;
                                       \nonumber\\
& & \Phi = \Phi_{e}-\frac{\phi}{\beta} +\mathcal{O}\left(
\beta^{-2}\right)\,,
 \qquad \Phi_e=\frac{Q_e M_e}{2\sqrt{J_e^2+Q_e^4/4}} \,,
 \qquad \phi=\frac{\pi Q_e^3}{\sqrt{J_e^2+Q_e^4/4}}\,.
   \nonumber  \label{KN:expand}
\eea
The extremal entropy $S_e$, and conjugated potentials $\omega$ and
$\phi$ agree with the corresponding quantities computed in
\cite{Astefanesei:2006dd} using Sen's entropy function
formalism\footnote{The match is obtained once the different
conventions for the action and $G_4$ are mapped. In
\cite{Astefanesei:2006dd}, the Maxwell term in the action has an
extra factor of $1/4$ and $G_4\equiv 1/(16\pi)$.}.

The QSR for this system is
\bea && I= \beta\left( M_e-\Phi_{e}Q_e-\Omega_{e}J_{e}\right)+
\phi\, Q_e+\omega\, J_e-S_{e} +\mathcal{O}\left( \beta^{-1}\right)\,
 \label{KN:QSR}
\eea
The first term does not vanish, a feature that seems to be common to
non-BPS extreme black holes with ergosphere. The factor in between
brackets is $M_e(M_e^2-Q_e^2)/(M_e^2+a_e^2)$. If rotation is absent,
$a=0$, one has $M_e=Q_e$ and the above term vanishes. When it is
present, the solution has an ergosphere and the non-vanishing
contribution seems to be associated with its existence, as discussed
in section \ref{sec:Extreme}.


\end{document}